%% file: main.tex
\documentclass[lettersize,journal]{IEEEtran}

\usepackage{amsmath,amsfonts}
\usepackage{algorithmic}
\usepackage{algorithm}
\usepackage{array}
\usepackage[caption=false,font=normalsize,labelfont=sf,textfont=sf]{subfig}
\usepackage{textcomp}
\usepackage{stfloats}
\usepackage{url}
\usepackage{verbatim}
\usepackage{graphicx}
\usepackage{cite}
\input{scripts/0_meta}

\begin{document}

\title{\toolname{}: Agent-Based Root Cause Diagnosis for Kernel Crashes}

\author{
Weijing Wang,
Zan Wang,
Dong Wang,
Haichi Wang,
and Junjie Chen%

\thanks{
This work was supported by the National Key Research and Development Program of China (Grant No. 2024YFB4506300), and the National Natural Science Foundation of China (Grant Nos. 62322208, 62472310).
}
\thanks{
Weijing Wang, Dong Wang, Haichi Wang, and Junjie Chen are with the School of Computer Software, Tianjin University, Tianjin 300350, China. 
Zan Wang is with Tianjin University, Tianjin 300350, China.
E-mail: \{\nolinkurl{wangweijing, wangzan, dong_w, wanghaichi, junjiechen}\}@tju.edu.cn.
}
\thanks{
Corresponding authors: Zan Wang and Dong Wang.
}
}








\maketitle

\begin{abstract}
The Linux kernel is one of the most complex software systems, where automated fuzzing continuously exposes thousands of crashes, yet root-cause diagnosis remains a manual and time-consuming bottleneck.
Existing LLM-based root cause analysis (RCA) techniques, effective for distributed systems, do not readily generalize to kernel debugging due to sparse low-level artifacts, heterogeneous diagnostic evidence (e.g., syscalls, logs, and crash reports), and complex non-linear fault propagation that demands fine-grained method-level reasoning.
To address these challenges, we propose \toolname{}, an agent-based framework for kernel root-cause diagnosis via structured causal reasoning.
\toolname{} first aligns heterogeneous diagnostic artifacts through log-to-code mapping, and then employs artifact-specialized agents to iteratively reason over source-level program semantics and crash-specific kernel configurations. The inferred causal dependencies are incrementally organized into structured Evidence Graphs, enabling accurate faulty-method localization and causal explanations.
We evaluate \toolname{} on the real-world KGYM benchmark. 
\toolname{} consistently outperforms state-of-the-art localization approaches at both file and method levels, achieving significant improvements in Top@k accuracy, including up to 4× and 2× gains in challenging settings without explicit hints. 
Furthermore, both human and LLM-assisted evaluations show that \toolname{} generates accurate, coherent, and actionable diagnostic explanations.
Overall, this work lays the foundation for automated kernel root-cause diagnosis by bridging low-level diagnostic evidence with source-level causal reasoning.
\end{abstract}


\input{scripts/1_introduction_v2}

\input{scripts/2_motivation_v3}
\input{scripts/3_approach_v3}

\input{scripts/4_evaluation_v2}

\input{scripts/5_rq1_v3}

\input{scripts/6_discussion}

\input{scripts/7_related_work}

\input{scripts/8_conclusion}


 
%



\bibliographystyle{IEEEtran}
\bibliography{reference}

\end{document}

%% file: scripts/0_meta.tex
\usepackage{booktabs}
\usepackage{multirow}
\usepackage{pifont}
\usepackage{url}
\usepackage{xcolor}
\usepackage{hyperref}
\usepackage{graphicx}

\usepackage{textcomp}
\usepackage[T1]{fontenc}
\usepackage{xcolor}
\usepackage{listings}
\usepackage[most]{tcolorbox}
\usepackage{newfloat}
\usepackage{enumitem}

\usepackage{tabularx}
\usepackage{colortbl} 

\usepackage{amsmath}

\usepackage[most]{tcolorbox}

\newcommand{\finding}[2]{
  \begin{tcolorbox}[
    colback=gray!10,   
    colframe=gray!70, 
    boxrule=0pt,       
    borderline west={2pt}{0pt}{gray}, 
    sharp corners,      
    enhanced
  ]
  \textbf{Answering #1:} #2
  \end{tcolorbox}
}

\lstdefinestyle{compactcode}{
language=C,
upquote=true,             
basicstyle=\ttfamily\fontsize{7.2pt}{8.0pt}\selectfont,
keywordstyle=\color{blue!70!black}\bfseries,
commentstyle=\color{green!40!black}\itshape,
stringstyle=\color{orange!70!black},
numbers=none,
breaklines=true, 
breakatwhitespace=false, 
showstringspaces=false, 
tabsize=2,      
frame=none,
aboveskip=0pt,
belowskip=0pt,
columns=fullflexible,
keepspaces=true
}

\newtcblisting{fullfilecode}[2][]{
enhanced,
arc=1.0mm,
outer arc=1.0mm,
boxrule=0.4pt,
colframe=black!18,
colback=black!1,
colbacktitle=gray!12,
coltitle=black,
fonttitle=\ttfamily\fontsize{7.2pt}{8.0pt}\selectfont\bfseries,
title={#2},
toptitle=0.6mm, 
bottomtitle=0.6mm,
left=0.8mm,
right=0.8mm,
top=0.4mm,
bottom=0.4mm,
boxsep=0pt, 
listing only,
listing options={style=compactcode},
#1
}

\lstdefinestyle{reportlog}{
  upquote=true,             
  basicstyle=\ttfamily\fontsize{6.8pt}{7.8pt}\selectfont, 
  columns=fullflexible,
  breaklines=true,
  breakatwhitespace=false,
  showstringspaces=false,
  keepspaces=true,
  numbers=none,
  keywordstyle=\color{black}, 
  commentstyle=\color{black},
  stringstyle=\color{black},
  emph={ERROR, WARNING, CRITICAL, BUG, failed},
  emphstyle=\bfseries\color{red!70!black},
  extendedchars=true,
  literate={_}{{\textunderscore}}1 {~}{{\textasciitilde}}1, 
  escapeinside={(*@}{@*)},
}

\definecolor{myblue}{HTML}{B5D5FF}

\newtcolorbox{textbox}[2][]{
  enhanced,
  arc=0.5mm,
  outer arc=0.5mm,
  boxrule=0.5pt,
  colframe=myblue!40!gray,
  colback=myblue!10!white,
  colbacktitle=myblue!20!,
  coltitle=black,
  fonttitle=\ttfamily\fontsize{7.5pt}{8.5pt}\selectfont\bfseries,
  title={#2},
  toptitle=0.6mm,
  bottomtitle=0.6mm,
  left=1mm,
  right=1mm,
  top=0.8mm,
  bottom=0.8mm,
  boxsep=0.5mm,
  before upper={\parindent0pt},
  fontupper=\fontsize{7pt}{7.8pt}\selectfont, 
  #1
}

\newtcblisting{reportbox}[2][]{
  enhanced,
  arc=0.5mm,
  outer arc=0.5mm,
  boxrule=0.5pt,
  colframe=purple!30!gray, 
  colback=purple!2!white, 
  colbacktitle=purple!10!,
  coltitle=black,
  fonttitle=\ttfamily\fontsize{7.5pt}{8.5pt}\selectfont\bfseries,
  title={#2},
  toptitle=0.6mm,
  bottomtitle=0.6mm,
  left=1mm,
  right=1mm,
  top=0.5mm,
  bottom=0.5mm,
  boxsep=0pt, 
  listing only,
  listing options={style=reportlog},
  #1
}

\newtcblisting{logbox}[2][]{
  enhanced,
  arc=0.5mm,
  outer arc=0.5mm,
  boxrule=0.5pt,
  colframe=green!25!gray,       
  colback=green!2!white,        
  colbacktitle=green!10!,
  coltitle=black,
  fonttitle=\ttfamily\fontsize{7.5pt}{8.5pt}\selectfont\bfseries,
  title={#2},
  toptitle=0.6mm,
  bottomtitle=0.6mm,
  left=1mm,
  right=1mm,
  top=0.5mm,
  bottom=0.5mm,
  boxsep=0pt, 
  listing only,
  listing options={style=reportlog},
  #1
}

\newtcblisting{syscallbox}[2][]{
  enhanced,
  arc=0.5mm,
  outer arc=0.5mm,
  boxrule=0.5pt,
  colframe=blue!20!gray,       
  colback=blue!2!white,        
  colbacktitle=blue!10!,
  coltitle=black,
  fonttitle=\ttfamily\fontsize{7.5pt}{8.5pt}\selectfont\bfseries,
  title={#2},
  toptitle=0.6mm,
  bottomtitle=0.6mm,
  left=1mm,
  right=1mm,
  top=0.5mm,
  bottom=0.5mm,
  boxsep=0pt, 
  listing only,
  listing options={style=reportlog},
  #1
}
\newcommand{\logln}[1]{%
  \makebox[2.8em][r]{\textcolor{gray}{\scriptsize #1}\hspace{0.8em}}%
}

\definecolor{schemaOperation}{HTML}{6A3D9A} 
\definecolor{schemaStructure}{HTML}{175A8A} 
\definecolor{schemaLocation}{HTML}{B45309} 
\definecolor{schemaEvidence}{HTML}{2F6B4F} 
\definecolor{schemaString}{HTML}{4F5965} 
\definecolor{schemaNumber}{HTML}{4F5965} 

\lstdefinelanguage{EvidenceSchema}{
sensitive=true,
alsoletter={_},
morekeywords=[1]{add_node,add_edge},
morekeywords=[2]{id,type,src,dst},
morekeywords=[3]{loc_file,loc_method,loc_line},
morekeywords=[4]{evidence},
morestring=[b]{'},
literate=*
{0}{{{\color{schemaNumber}0}}}1
{1}{{{\color{schemaNumber}1}}}1
{2}{{{\color{schemaNumber}2}}}1
{3}{{{\color{schemaNumber}3}}}1
{4}{{{\color{schemaNumber}4}}}1
{5}{{{\color{schemaNumber}5}}}1
{6}{{{\color{schemaNumber}6}}}1
{7}{{{\color{schemaNumber}7}}}1
{8}{{{\color{schemaNumber}8}}}1
{9}{{{\color{schemaNumber}9}}}1
}

\lstdefinestyle{evidenceschema}{
language=EvidenceSchema,
upquote=true,
basicstyle=\ttfamily\fontsize{6.8pt}{7.8pt}\selectfont,
columns=fullflexible,
keepspaces=true,
breaklines=true,
breakatwhitespace=false,
showstringspaces=false,
numbers=none,
tabsize=2,
identifierstyle=\color{black},
keywordstyle=[1]\bfseries\color{schemaOperation},
keywordstyle=[2]\bfseries\color{schemaStructure},
keywordstyle=[3]\bfseries\color{schemaLocation},
keywordstyle=[4]\bfseries\color{schemaEvidence},
stringstyle=\color{schemaString},
}

\newtcblisting{reportgraphbox}[2][]{
  enhanced,
  arc=0.5mm,
  outer arc=0.5mm,
  boxrule=0.5pt,
  colframe=purple!30!gray, 
  colback=purple!2!white, 
  colbacktitle=purple!10!,
  coltitle=black,
  fonttitle=\ttfamily\fontsize{7.5pt}{8.5pt}\selectfont\bfseries,
  title={#2},
  toptitle=0.6mm,
  bottomtitle=0.6mm,
  left=1mm,
  right=1mm,
  top=0.5mm,
  bottom=0.5mm,
  boxsep=0pt, 
  listing only,
  listing options={style=evidenceschema},
  #1
}

\newtcblisting{loggraphbox}[2][]{
  enhanced,
  arc=0.5mm,
  outer arc=0.5mm,
  boxrule=0.5pt,
  colframe=green!25!gray,       
  colback=green!2!white,        
  colbacktitle=green!10!,
  coltitle=black,
  fonttitle=\ttfamily\fontsize{7.5pt}{8.5pt}\selectfont\bfseries,
  title={#2},
  toptitle=0.6mm,
  bottomtitle=0.6mm,
  left=1mm,
  right=1mm,
  top=0.5mm,
  bottom=0.5mm,
  boxsep=0pt, 
  listing only,
  listing options={style=evidenceschema},
  #1
}

\newtcblisting{syscallgraphbox}[2][]{
  enhanced,
  arc=0.5mm,
  outer arc=0.5mm,
  boxrule=0.5pt,
  colframe=blue!20!gray,       
  colback=blue!2!white,        
  colbacktitle=blue!10!,
  coltitle=black,
  fonttitle=\ttfamily\fontsize{7.5pt}{8.5pt}\selectfont\bfseries,
  title={#2},
  toptitle=0.6mm,
  bottomtitle=0.6mm,
  left=1mm,
  right=1mm,
  top=0.5mm,
  bottom=0.5mm,
  boxsep=0pt, 
  listing only,
  listing options={style=evidenceschema},
  #1
}

\newcommand{\toolname}{KernelDiag}

\newcommand{\agentless}{\textit{Agentless}}
\newcommand{\linuxfl}{\textit{LinuxFL$^+$}}
\newcommand{\agentlessDiag}{\textit{Agentless}-Diag}
\newcommand{\linuxflDiag}{\textit{LinuxFL$^+$}-Diag}

\newcommand{\agentlessnormal}{Agentless}
\newcommand{\linuxflnormal}{LinuxFL$^+$}

\newcommand{\toolwosyscall}{\textit{w/o} Syscall}

\newcommand{\toolwolog}{\textit{w/o} Log}

\newcommand{\toolworeport}{\textit{w/o} Report}

\newcommand{\toolnoagent}{\textit{w/} DirectQuery}

\newcommand{\toolsingleagent}{\textit{w/} SingleAgent}

\newcommand{\toolnomapping}{\textit{w/o} Log-to-CodeMapping}
\newcommand{\toolnoconfig}{\textit{w/o} EnvironmentPruning}
\newcommand{\toolnointrospection}{\textit{w/o} FunctionIntrospection}
\newcommand{\toolnograph}{\textit{w/o} EvidenceGraph}

\newcommand{\datafile}{\textit{F-All}}
\newcommand{\datafilehint}{\textit{F-Hint}}
\newcommand{\datafilenohint}{\textit{F-NoHint}}

\newcommand{\datamethod}{\textit{M-All}}
\newcommand{\datamethodhint}{\textit{M-Hint}}
\newcommand{\datamethodnohint}{\textit{M-NoHint}}

\definecolor{groupC}{HTML}{003153} 
\definecolor{groupB}{HTML}{7A4171} 
\definecolor{groupA}{HTML}{8B4513} 


%% file: scripts/1_introduction_v2.tex
\section{Introduction}

The Linux kernel underpins modern computing infrastructure, powering cloud platforms, mobile devices, and high-performance computing clusters~\cite{linux_kernel_org,wang2025empirical}. 
With over 40 million lines of code and a relentless cycle of continuous integration, it represents one of the most complex software systems~\cite{linux_kernel_40m}. 
To maintain its reliability, the community has deployed several testing and fuzzing frameworks like \texttt{Syzkaller} \cite{syzkaller} and \texttt{Syzbot}~\cite{syzbot}, which uncover thousands of previously unknown bugs each year.
Despite these advances in automated failure discovery, diagnosing these crashes remains a labor-intensive and manual bottleneck for developers~\cite{li2025yesterday, lin2022grebe}.

Root cause diagnosis aims to identify the specific trigger of a system failure and elucidate the propagation logic that connects the initial fault to the observed manifestation, thereby supporting developers in debugging software systems~\cite{li2022causal, xin2023causalrca,chen2024automatic}.
Recently, Root Cause Analysis (RCA) has seen substantial progress in cloud services and distributed systems, largely driven by the integration of Large Language Models (LLMs)~\cite{yao2022react,roy2024exploring,wang2024rcagent,xu2025openrca}.
Emerging LLM-based RCA approaches demonstrate strong capabilities in summarizing failure contexts and generating plausible explanations by leveraging semantic reasoning over observability data. 
These approaches typically synthesize high-level signals collected at runtime, such as distributed traces, application logs, and performance metrics, into coherent diagnostic insights, and have shown promising results in industrial scenarios.

Nevertheless, these existing RCA approaches do not readily transfer to the Linux kernel for two fundamental reasons.
First, the rich operational context (such as end-to-end request traces and structured logs) available in cloud environments is largely absent in kernel scenarios~\cite{mathai2024kgym, zhou2025benchmarking}.
In contrast, kernel debugging typically relies on low-level and sparse artifacts, which provide limited semantic context and fragmented execution information.
Second, existing RCA approaches often settle for coarse-grained outputs, such as 
classifying failure type~\cite{zhang2025scalalog, sui2023logkg, zhang2024multivariate}
or identifying faulty modules, services, or components~\cite {li2021practical, liu2021microhecl, xie2023unsupervised, zeng2023traceark,yu2023nezha}, which is often sufficient in distributed systems for rapid mitigation.
In the Linux kernel, however, the complexity of the codebase demands fine-grained, method-level localization to support actionable debugging and patching.
Without such precision, developers are left with a significant root cause--manifestation gap, where reconstructing fault propagation across complex kernel subsystems remains largely manual and error-prone.

Recent work has begun to explore kernel fault localization, such as \linuxfl{}~\cite{zhou2025benchmarking}.
However, its effectiveness is limited by its reliance on crash reports as the primary diagnostic signal. 
When the root cause is not explicitly reflected in the report, like when the fault originates from earlier execution steps or involves complex propagation paths, its performance degrades significantly. 
Moreover, \linuxfl{} focuses solely on identifying faulty locations, without explaining how the fault propagates to the crash, thereby providing developers with an incomplete and less actionable diagnosis.
These limitations suggest that crash reports alone are insufficient for in-depth diagnosis. 
Instead, kernel failures are often accompanied by multiple complementary artifacts, including triggering syscalls, runtime logs, and crash reports, each capturing different aspects of execution, from external inputs to internal states and failure manifestations.
Effectively leveraging these heterogeneous artifacts, however, remains challenging:\\
\textbf{\textit{Challenge I:  Aligning and interpreting heterogeneous multi-source artifacts.}}
Kernel diagnostic artifacts are inherently heterogeneous in both structure and abstraction level. Triggering syscalls describe external inputs at the user–kernel boundary, runtime logs capture partial internal states across subsystems, and crash reports reflect failure manifestations at the point of failure (e.g., stack traces and error contexts). These artifacts are often loosely coupled, temporally misaligned, and expressed in different formats, resulting in fragmented and partially overlapping execution views. Without explicit alignment and normalization, it is difficult to establish a common semantic space and recover a coherent execution context from distributed evidence.\\
\textbf{\textit{Challenge II: Reasoning about complex and non-linear root-cause propagation.}}
Even with aligned evidence, reconstructing how faults propagate through the kernel remains highly non-trivial. The Linux kernel exhibits intricate execution semantics, including indirect control flows, macro expansions, interrupt-driven and asynchronous callbacks, and configuration-dependent behaviors. These characteristics lead to non-linear, discontinuous, and sometimes implicit execution paths that are not explicitly captured in any single artifact. Consequently, tracing causal dependencies across multiple functions, subsystems, and execution stages requires sophisticated reasoning over incomplete and distributed evidence, making root-cause propagation analysis inherently difficult.

To address the aforementioned challenges, 
we propose \textit{\toolname{}}, an agent-based framework for kernel root-cause diagnosis built on structured heterogeneous evidence reasoning.
\begin{itemize}
    \item To address \textbf{\textit{Challenge I}}, \toolname{} first aligns heterogeneous artifacts into a common source-level semantic space through static-analysis-based log-to-code mapping.
    By expanding kernel logging macros and resolving log-emitting functions, it maps unstructured logs to precise code locations and recovers fragmented execution context, enabling cross-artifact correlation at the source-code level.
    Based on the aligned artifacts, \toolname{} decomposes heterogeneous evidence interpretation through artifact-specialized agents, each responsible for analyzing a specific evidence source, including triggering syscalls, runtime logs, and crash reports. 
    This design preserves source-specific semantics while reducing the complexity of cross-artifact reasoning.
    \item To address \textbf{\textit{Challenge II}}, \toolname{} augments iterative agent reasoning with kernel-aware command-line tools for in-depth semantic analysis of code. 
    Specifically, \toolname{} employs \textit{Semantic Function Introspection} to inspect full function bodies, branch conditions, error-handling logic, and state-update logic, enabling agents to uncover implicit causal dependencies that are not directly visible in any single artifact. 
    Moreover, \toolname{} applies \textit{Environment-Aware Semantic Pruning} to resolve pointer-referenced structures to their concrete state owners and eliminate semantically infeasible execution paths under runtime configurations.
    Together, these techniques enable reasoning over indirect, non-linear, and configuration-dependent fault propagation paths.
\end{itemize}
Through iterative exploration, each agent incrementally incorporates discovered evidence and inferred causal dependencies into a structured
\textit{Evidence Graph}, 
whose nodes represent source-level kernel entities and record their source locations to support cross-artifact alignment, 
while its edges capture inferred propagation relationships.
Leveraging these graphs, \toolname{} performs coarse-to-fine root-cause localization to identify the faulty method, followed by agent-driven root-cause explanation that traces the observed failure symptoms to the underlying logic flaw.

To evaluate the effectiveness of \toolname{}, we conduct a comprehensive study on the kernel crash dataset KGYM~\cite{mathai2024kgym}. 
For root-cause localization, we compare \toolname{} with state-of-the-art fault localization approaches, including the kernel-specific \linuxfl{}~\cite{zhou2025benchmarking} and the general-purpose \agentless{}~\cite{xia2024agentless}, at both the file and method levels. \toolname{} consistently outperforms all baselines, achieving 65.95\% and 33.33\% Top@1 accuracy at the file and method levels, improving over the strongest baseline by 27.78\% and 25.68\%, respectively. The advantage becomes even more pronounced on challenging cases without explicit root-cause hints, where \toolname{} achieves up to 4× and 2× higher Top@10 accuracy at the file and method levels, respectively.
For root-cause explanation, we conduct both LLM-assisted and human evaluations. 
The results demonstrate that \toolname{} consistently produces more accurate, coherent, and actionable diagnostic explanations than competing approaches.
On an additional post-release benchmark beyond the LLM's pre-training horizon, \toolname{} consistently outperforms existing approaches, improving Top@1 localization by up to 20.69\% at the file level and 60.00\% at the method level while maintaining superior explanation quality.

\smallskip
\noindent
\textbf{Contributions.} To sum up, our work makes the following major contributions: 
\textbf{(I)} We propose \toolname{}, a novel agent-based framework for kernel root-cause diagnosis that leverages structured reasoning over heterogeneous artifacts to model fault propagation.
\textbf{(II)} \toolname{} demonstrates superior root-cause localization effectiveness on both file and method levels, especially in challenging settings without explicit hints.
\textbf{(III)} \toolname{} generates high-quality diagnostic explanations, potentially supporting developers in understanding and fixing kernel failures. 
To facilitate future research and reproducibility, we publicly release the replication package~\cite{kerneldiag2026}.

%% file: scripts/2_motivation_v3.tex
\section{Motivation}
\label{ref:motivation}

We now present a motivating example to demonstrate our key insight: effective kernel crash diagnosis requires leveraging complementary artifacts to enable cross-source reasoning.
Specifically, Figure~\ref{fig:clue_in_log} illustrates a real-world EXT4 filesystem crash\footnote{https://syzkaller.appspot.com/bug?extid=c7358a3cd05ee786eb31} reported by the Syzbot system.

\input{figures/important_log}

When a kernel crash occurs, developers typically begin diagnosis from the \textbf{crash report}, treating it as the primary diagnostic signal.
Unlike issue reports in distributed systems that provide rich contextual descriptions and anomalous logs~\cite{li2025coca,chen2020towards}, kernel crash reports expose only the low-level terminal execution state.
As a result, they often fail to reveal the true root cause. 
Given the motivating example, the ground truth
indicates that
\textit{\_\_ext4\_fill\_super} fails to validate the metadata state, allows 
a corrupted superblock to persist, 
and eventually triggers a \texttt{BUG()} in \textit{ext4\_es\_cache\_extent}.
However, the crash report in Figure~\ref{fig:clue_in_log}(a) neither mentions the ground-truth buggy file (\textit{fs/ext4/super.c}) nor the faulty method (\textit{\_\_ext4\_fill\_super}), and provides no insight into this underlying failure mechanism.
Consequently, models relying primarily on crash traces are prone to incomplete or misleading reasoning.

Beyond crash reports, two additional artifacts play complementary roles in kernel crash diagnosis: \textbf{syscalls} and \textbf{runtime logs}.
Syscalls capture the high-level operational intent at the user–kernel interface~\cite{hao2023syzdescribe}, providing a structured view of how the failure is triggered.
In contrast, runtime logs record unstructured observations of internal states across kernel subsystems, offering relatively fine-grained clues about anomalies during execution.
In this example, the triggering syscall \texttt{syz\_mount\_image\$ext4} in Figure~\ref{fig:clue_in_log}(b) aims to mount an EXT4 disk image and initialize the corresponding filesystem metadata in the kernel.
During the execution, it triggers an extensive chain of functions in \textit{fs/ext4/}, including \textit{ext4\_get\_tree} for mount setup, \textit{ext4\_fill\_super} and \textit{\_\_ext4\_fill\_super} for superblock initialization and metadata loading, \textit{ext4\_check\_descriptors} for descriptor validation, 
and other extent- and directory-processing routines, each of which may appear to be a plausible fault method from the developers' perspective.
Meanwhile, the runtime logs in Figure~\ref{fig:clue_in_log}(c) expose a critical clue at line 651, a \textit{checksum failure} in \textit{ext4\_check\_descriptors} called by \textit{\_\_ext4\_fill\_super}, which signals an early metadata inconsistency. 
Thereby, the log anomaly narrows the likely culprit to the mount-time metadata handling logic around \textit{\_\_ext4\_fill\_super}, rather than higher-level wrapper functions such as \textit{ext4\_get\_tree} or \textit{ext4\_fill\_super}.
By correlating this log-based anomaly with the syscall-derived sequence of EXT4 routines, we can reconstruct the complete failure propagation: \textit{\_\_ext4\_fill\_super} fails to properly validate and reject corrupted filesystem metadata during the mounting phase. 
This allows the latent inconsistency to persist and propagate through subsequent filesystem operations, ultimately triggering the terminal \texttt{BUG()} crash within \textit{ext4\_es\_cache\_extent}.

This example highlights that effective kernel crash diagnosis requires joint reasoning over heterogeneous artifacts. However, such reasoning is inherently challenging. These artifacts differ substantially in abstraction, structure, and granularity, ranging from structured execution traces to unstructured natural-language logs, making unified alignment and reasoning difficult. Moreover, kernel failures often propagate through long, non-linear execution paths due to the kernel's complex nature, where the root cause is distant from the observed crash site.
These challenges together lead to fragmented evidence and obscured causal chains, making it difficult to accurately align cross-source signals and precisely pinpoint the true root cause in the kernel context.

%% file: figures/important_log.tex
\begin{figure}[ht!]
\centering
\begin{minipage}{0.99\columnwidth}
\begin{reportbox}{(a) Crash Report: kernel BUG in ext4\_es\_cache\_extent}
kernel BUG at fs/ext4/extents_status.c:899!
...
RSP: 0018:ffffc90001abee80 EFLAGS: 00010293
RAX: 0000000000000000 RBX: 00000000000013de RCX: 00000000000000 ...
Call Trace:
 <TASK>
 ext4_cache_extents+0x13e/0x2d0 fs/ext4/extents.c:518
 ext4_find_extent+0x8f6/0xd10 fs/ext4/extents.c:916
 ext4_ext_map_blocks+0x1e2/0x5f30 fs/ext4/extents.c:4098
 ext4_map_blocks+0x9ca/0x18a0 fs/ext4/inode.c:563
 ext4_getblk+0x553/0x6b0 fs/ext4/inode.c:849
 ext4_bread_batch+0x7c/0x550 fs/ext4/inode.c:923
 __ext4_find_entry+0x482/0x1050 fs/ext4/namei.c:1600
 ext4_lookup_entry fs/ext4/namei.c:1701 [inline]
 ext4_lookup fs/ext4/namei.c:1769 [inline]
 ext4_lookup+0x4fc/0x730 fs/ext4/namei.c:1760
 __lookup_slow+0x24c/0x480 fs/namei.c:1707
 ...
RIP: 0033:0x7f8e44f22f19 ...

\end{reportbox}

\vspace{-2.5mm}
\begin{syscallbox}{(b) Syscall}
r0 = syz_mount_image$ext4(...); openat(r0, ...); 
\end{syscallbox}
\vspace{-2.5mm}
\begin{logbox}{(c) Partial Logs}
(*@\logln{1}@*)cgroup: Unknown subsys name 'net'
(*@\logln{647}@*)loop0: detected capacity change from 0 to 1051
(*@\logln{651}@*)EXT4-fs (loop0): ext4_check_descriptors: Checksum for group 0 failed (60935!=0)
(*@\logln{654}@*)EXT4-fs (loop0): mounted filesystem without journal. Quota mode: writeback.
\end{logbox}

\end{minipage}
\caption{An example of a real-world kernel crash}
\label{fig:clue_in_log}
\vspace{-1em}
\end{figure}

%% file: scripts/3_approach_v3.tex
\section{Approach}

\begin{figure*}[htbp!]
    \centering
    \includegraphics[width=0.98\linewidth]{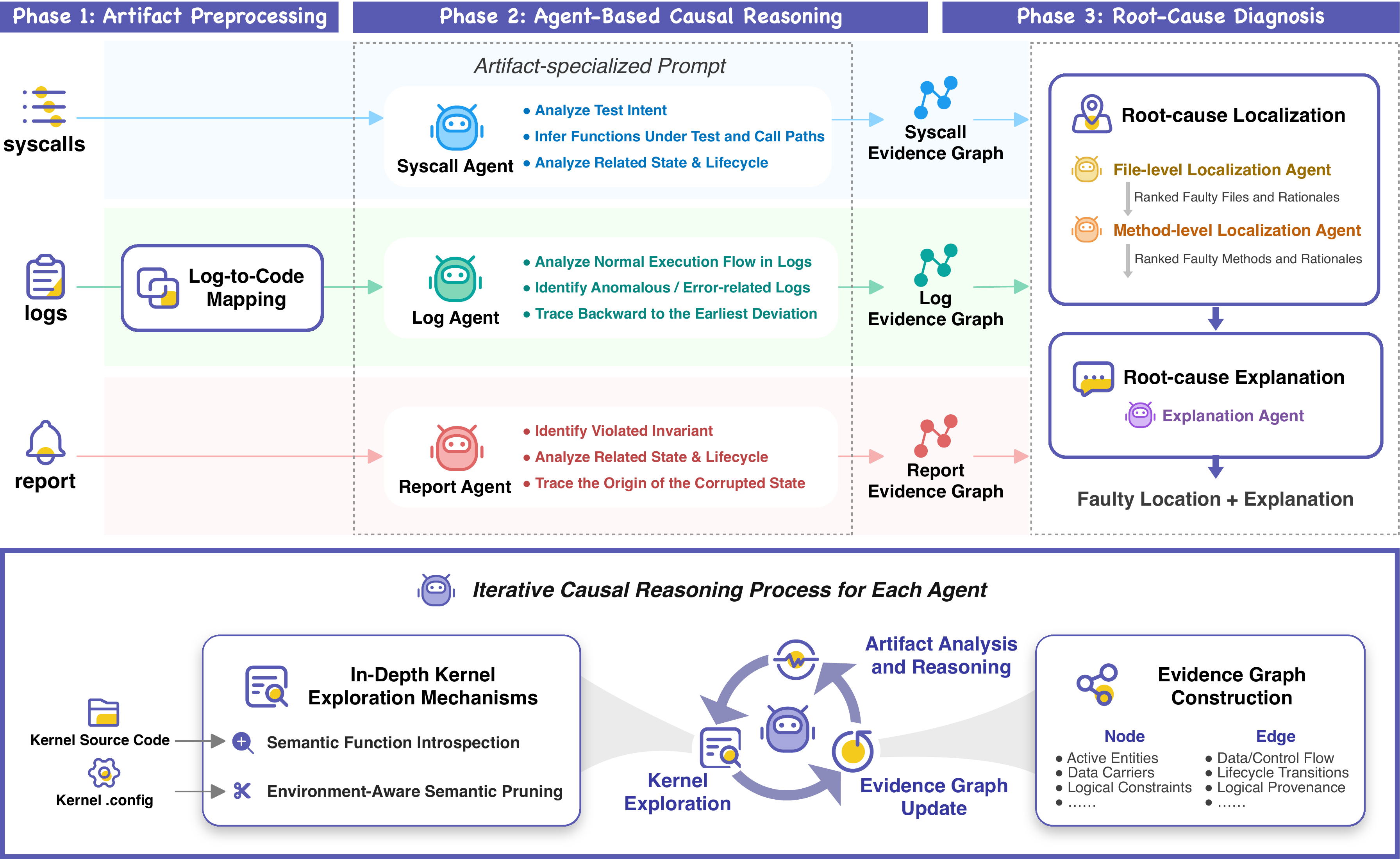}
    \caption{Overview of \toolname{}}
        \label{fig:overview}
        \vspace{-0.5em}
\end{figure*}

In this paper, we propose \toolname{}, an agent-based framework for kernel root-cause diagnosis built on structured reasoning over heterogeneous evidence.
Figure~\ref{fig:overview} presents the overall architecture of \toolname{}. 
First, \toolname{} performs static-analysis-based log-to-code mapping to align runtime log evidence with source code (Section \ref{sec:log-to-code-mapping}). 
In the causal reasoning phase, \toolname{} first interprets heterogeneous artifacts through role-specialized agent-based analysis (Section \ref{sec:multi-agent}-I).
Each artifact-specific agent then conducts in-depth kernel code investigation via Semantic Function Introspection and Environment-Aware Semantic Pruning (Section \ref{sec:multi-agent}-II), and constructs a topologically structured Evidence Graph to capture root-cause propagation (Section \ref{sec:multi-agent}-III). 
Finally, in the root-cause diagnosis phase, \toolname{} performs agent-driven coarse-to-fine root-cause localization based on the Evidence Graphs produced by artifact-specific agents (Section \ref{sec:root-cause-diagnosis}-I), 
and generates a causal explanation for the target method (Section \ref{sec:root-cause-diagnosis}-II).

\subsection{Log-to-Code Mapping}
\label{sec:log-to-code-mapping}

Among heterogeneous artifacts, runtime logs often contain critical clues, but their lack of explicit source-level grounding 
hinders effective integration
with other artifacts such as crash reports and syscalls.
This necessitates accurate log-to-code mapping to anchor log evidence to concrete program locations.
While some entries (e.g., line 651 in Figure~\ref{fig:clue_in_log}) expose function names such as \textit{ext4\_check\_descriptors}, most logs lack explicit source-level identifiers, leaving their origins ambiguous.
Existing approaches typically identify logging statements through keyword matching (e.g., ``log'', ``print'',  and ``info'') or pre-defined API rules~\cite{li2025coca,shu2025empirical,huo2023autolog}, assuming logging functions are limited, stable, and explicitly identifiable. 
These assumptions break in the Linux kernel. Logging is decentralized and heavily macro-encapsulated, with subsystem-specific wrappers obscuring logging intent~\cite{patel2022sense}.
Figure~\ref{fig:example1} illustrates the complexity of kernel logging, where logging behavior is hidden behind multiple layers of macro abstraction. 
The example centers on \texttt{invalfc}, which produces the first log line in the aforementioned crash example (Figure~\ref{fig:clue_in_log}(c)).
Figure~\ref{fig:example1}(a) shows the call site where logging is triggered but not explicit, (b) presents the macro expansion chain that obscures the logging semantics, and (c) shows the underlying function definition where \texttt{printk} is finally invoked. 
Notably, \texttt{invalfc} neither contains logging-related keywords nor directly calls \texttt{printk}, making its intent difficult to identify.

\input{figures/example_1_logging}

To align fragmented log evidence, \toolname{} introduces a \textbf{Log-to-Code Mapping} grounded in reachability analysis over logging abstractions.
Following the official Linux logging standards\footnote{https://docs.kernel.org/core-api/printk-index.html}, 
which encourage subsystem-specific wrappers over core logging primitives, 
\toolname{} formalizes logging identification as a backward reachability problem from atomic logging primitives, rather than relying on surface-level identifiers.
Specifically, \toolname{} first identifies a set of \textit{atomic logging sinks}, defined as functions or macros whose bodies (or fully expanded forms) directly invoke kernel logging primitives such as \texttt{printk}.
Building on these sinks, 
\toolname{} constructs a \textit{logging dependency flow} that captures both function-call and macro-expansion dependencies through static analysis of kernel source code with preprocessing-aware macro tracking.
\toolname{} then performs a backward transitive closure over the logging dependency flow from atomic sinks, propagating logging semantics across abstraction layers. 
As a result, any function or macro that reaches an atomic sink is treated as logging-relevant, even without explicit logging primitives or keywords.
In the example, \texttt{logfc} is first identified as an atomic logging sink 
because it directly invokes \texttt{printk}.
\toolname{} then performs backward reachability over the dependency flow and recovers the dependency chain:
$\texttt{logfc} \rightarrow \texttt{\_\_plog} \rightarrow \texttt{errorfc} \rightarrow \texttt{invalfc}$.
Through this transitive propagation, \texttt{invalfc}, despite lacking explicit logging semantics, is regarded as a logging-relevant macro. 
Consequently, its call site in Figure~\ref{fig:example1}(a) can be precisely mapped to the emitted log message.
This mechanism enables \toolname{} to ground  unstructured and ambiguous log evidence to concrete source-level locations, thereby bridging the gap between runtime observations and program structure.

\subsection{Agent-Based Causal Reasoning}
\label{sec:multi-agent}
During the causal reasoning, \toolname{} assigns each artifact to a role-specialized agent to construct an artifact-specific causal view of the crash. 
Given its input artifact, each agent iteratively explores the kernel codebase for the corresponding kernel version with predefined kernel-aware tools to reason about kernel-specific constructs such as wrappers, indirect dispatch, pointer-referenced states, and configuration-dependent execution paths, thereby reconstructing the fault-propagation chain.
Rather than producing free-form summaries, each agent generates and continuously updates a schema-validated Evidence Graph, whose nodes and edges encode source-grounded evidence and causal dependencies. 
These artifact-specific graphs share source-level location attributes, which serve as alignment anchors for merging heterogeneous evidence in the subsequent root-cause localization and explanation stages.

\smallskip
\noindent
\textbf{I. Agent-Based Design.}
To interpret heterogeneous artifacts, 
\toolname{} adopts a role-specialized agent-based framework that decomposes reasoning across artifact types. 
Instead of forcing a single model to jointly process heterogeneous inputs, 
each agent specializes in one artifact with its own reasoning directionality and causal anchor, enabling structured reasoning while reducing cross-artifact interference and cognitive overload.
We instantiate three distinct agents:
\begin{itemize}[leftmargin=15pt,nosep]
    \item \textbf{Syscall Agent:} Starting from user-space invocations, this agent analyzes syscall sequences, arguments, and invocation contexts to infer the functions under test, the kernel objects being manipulated, and the execution paths likely triggered inside the kernel. It further models the interaction between syscalls and kernel subsystems, enabling it to construct a coarse-grained execution skeleton. This skeleton serves as an operational baseline, providing contextual constraints for subsequent reasoning and narrowing the search space of potential fault locations.
    \item \textbf{Log Agent}: This agent treats runtime logs with source-level locations as observable execution events and interprets them as partial manifestations of internal state transitions.
    Upon detecting anomalous or error-related log entries, it performs backward reasoning along call relations, data dependencies, and object-state transitions to identify the earliest point of deviation from expected behavior. 
    Through this process, the agent uncovers fine-grained, lifecycle-level causal evidence that bridges the gap between execution and failure.
    \item \textbf{Report Agent}: 
    Starting from the crash site, this agent analyzes low-level failure signals such as panic messages, RIP locations, register states, and stack traces. 
    It identifies violated kernel invariants or API contracts (e.g., assertions and BUG\_ON conditions), and traces the origin of the corrupted state to reconstruct how these violations propagate through the execution stack. 
    By reasoning about control-flow and constraint violations, explains the terminal symptom and its immediate causes.
\end{itemize}

\smallskip
\noindent
\textbf{II. In-Depth Kernel Exploration Mechanisms.}

\begin{figure*}[t!]
    \centering
    \includegraphics[width=0.99\linewidth]{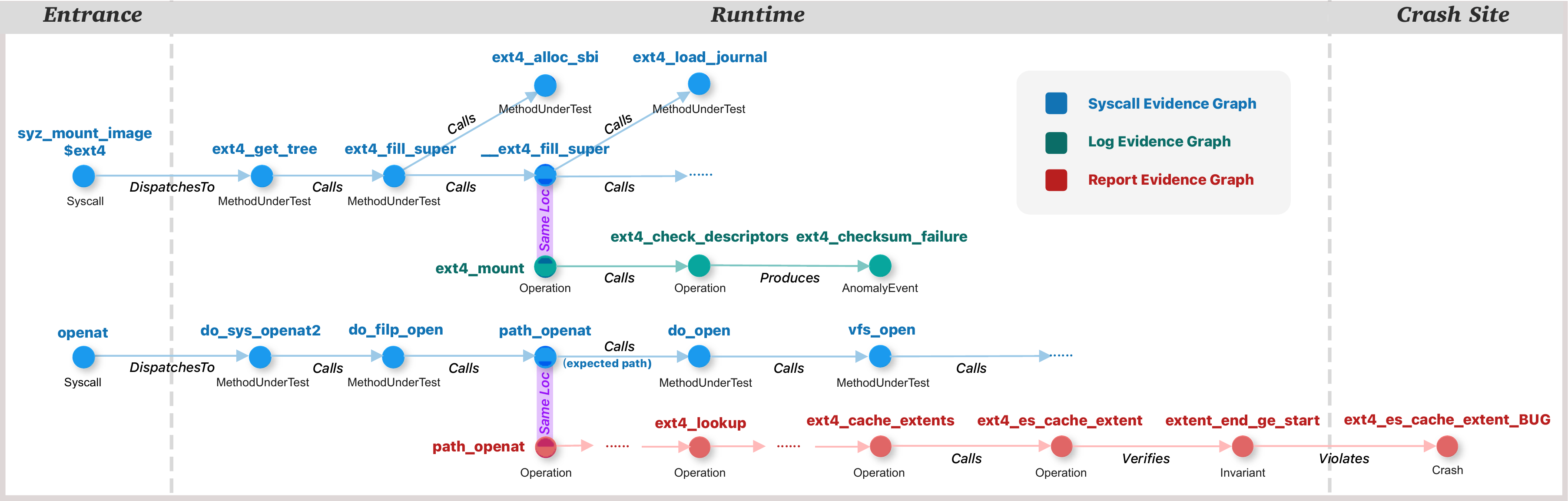}
    \caption{An example of causal topology structure of Evidence Graphs}
        \label{fig:graph_structure}
        \vspace{-0.5em}
\end{figure*}

While coarse AST-level summaries and call graphs provide useful structural guidance for repository exploration, they omit the fine-grained semantics needed for kernel crash diagnosis, including branch conditions, state validation, error handling, pointer-based updates, and configuration-dependent paths.
This limitation is amplified in the Linux kernel, where failures may propagate through macros, indirect dispatch, callbacks, and subsystem abstractions, making the causal path non-linear and only partially observable from static structure alone. 
To address it, \toolname{} equips each agent with kernel-aware command-line tools that go beyond structural navigation and support source-grounded semantic investigation:

\begin{itemize}[leftmargin=15pt,nosep]

    \item \textbf{Semantic Function Introspection}: 
    This mechanism equips agents with on-demand access to source-level semantic context beyond coarse structural navigation. 
    Prior to causal reasoning, \toolname{} performs a static indexing pass over the kernel codebase to construct a suite of kernel-aware tools that index the source locations and ranges of function definitions, macro definitions, and struct/union declarations.
    During exploration, agents can invoke these tools to retrieve the complete implementation of a target function, the definition of a macro, and the declaration of a relevant data structure according to their reasoning needs. 
    This allows agents to inspect branch conditions, error-handling logic, and state-update logic within complete function bodies, which are often missing from coarse AST-level summaries or call-graph-based exploration. 
    Moreover, \toolname{} performs structured field resolution for pointer-based accesses (e.g., \texttt{x->y}) by guiding agents to trace the accessed field through its struct definition to identify the corresponding kernel state owner and its related operations. 
    By connecting field accesses with the functions that initialize, validate, mutate, or consume the corresponding kernel object, agents can interpret state mutations and dependency relationships that are otherwise obscured by pointer indirection.
    \item \textbf{Environment-Aware Semantic Pruning}: This mechanism constrains agent exploration using the crash-specific kernel environment.
    In Linux, many execution paths are governed by build configurations, architectures, subsystems, and runtime object types, making seemingly reachable paths in the call graph infeasible for the observed execution.
    To reduce such ambiguity, \toolname{} enables agents to inspect environment artifacts, especially the kernel \texttt{.config} file, on demand. 
    When encountering configuration-dependent code, such as \texttt{\#ifdef CONFIG\_*}, \texttt{IS\_ENABLED(CONFIG\_*)}, architecture-specific branches, or subsystem-specific implementations, agents extract the corresponding conditions and verify whether they are active in the target environment.
    Based on this verification, agents eliminate infeasible branches, deprioritize inactive implementations, and focus subsequent analysis on source-level connected execution paths that are consistent with the observed environment.
    For indirect dispatch through callbacks, function pointers, or \texttt{ops} tables, agents further prioritize likely active handlers. 
    By pruning semantically infeasible paths early, this mechanism suppresses spurious exploration and grounds the resulting evidence in the actual kernel execution context.
\end{itemize}
Together, these mechanisms elevate agent reasoning from surface-level tracing to semantics-aware causal analysis. By grounding exploration in source-level semantics and crash-specific runtime context, agents can uncover latent dependencies and reconstruct complex fault propagation chains, enabling accurate root-cause diagnosis.

\smallskip
\noindent
\textbf{III. Causal Evidence Graph Construction.}
During exploration, \toolname{} guides each artifact-specific agent to incrementally record its findings as a schema-validated Evidence Graph, rather than free-form text, through a predefined command-line tool. 
In the Evidence Graph, \textbf{Nodes} represent kernel entities, including Active Entities (e.g., operations and anomalous events), Data Carriers (e.g., objects and buffers), and Logical Constraints (e.g., invariants and lifecycle states), while \textbf{Edges} capture causal or semantic dependencies such as Data/Control Flow, Lifecycle Transitions, and Logical Provenance. 
Although agents construct graphs from different artifacts, \toolname{} enforces a shared schema in which each \textit{Node} carries source-level location attributes, including file paths, function names, and line ranges, enabling cross-artifact entity alignment.
As a structured substrate for incremental reasoning, the Evidence Graph preserves causal topology across iterative analysis, enabling agents to reconstruct non-linear fault propagation chains in a globally consistent manner.
Figure~\ref{fig:graph_structure} presents the topology structure of evidence graphs from different artifacts according to the kernel crash shown in Figure~\ref{fig:clue_in_log} (Section~\ref{ref:motivation}).

\subsection{Root-Cause Diagnosis}
\label{sec:root-cause-diagnosis}

Root-cause diagnosis is an end-to-end task that identifies the faulty method and explains how its faulty logic propagates to the observed crash. 
Accordingly, \toolname{} first performs \textit{Root-cause Localization} based on the observed evidence, and then generates \textit{Root-cause Explanation} by pivoting on the identified method to reconstruct the fault-propagation process.

\smallskip
\noindent
\textbf{I. Root-cause Localization.}
Guided by unified evidence graphs constructed from heterogeneous artifacts, \toolname{}
identifies faulty methods through a two-stage, agent-driven process that prioritizes causal origin over crash proximity.

In the first stage, a \textbf{File-level Localization Agent} performs coarse-grained causal triage over Top-K suspicious files to establish a prior for fine-grained analysis. 
Rather than ranking files based on syntactic proximity to the crash site, the agent performs backward reasoning over the evidence graphs to identify the earliest point where abnormal behavior emerges,
such as invariant violations, broken lifecycle symmetry, or the introduction of invalid states. 
Based on this causal analysis, files are divided into  two tiers: Tier-1 files, which likely originate the fault, and Tier-2 files, which propagate, amplify, or expose the fault during execution.

In the second stage, a \textbf{Method-level Localization Agent} performs fine-grained inspection within and around the prioritized files to conduct the Top-K suspicious methods.
It examines both directly implicated methods and semantically related routines,
including producers, mutators, consumers, guards, and error handlers, to capture the full causal context. 
A method is identified as the root cause only when the agent provides an evidence-grounded explanation of how its logic introduces the faulty state and leads to the observed symptoms in the causal evidence graph.

Overall, this process reframes localization as a causality-driven inference task. 
Rather than relying on proximity-based heuristics, \toolname{} identifies root causes based on where faults originate, how they propagate, and whether the inferred mechanism is consistent with observed evidence, enabling more explainable localization.

\smallskip
\noindent
\textbf{II. Root-cause Explanation Generation.}
\toolname{} employs an \textbf{Explanation Agent} to generate the root-cause explanation. 
The agent takes as input the Evidence Graphs, the prior localization results and their rationales, and synthesizes a textual explanation centered on the target method. 
The root-cause explanation is required to be evidence-grounded and to cover two aspects: the concrete fault logic mechanism inside the target method, and the propagation process from the target method to the observed crash. 
Due to space constraints, the prompts used above are available in our public repository~\cite{kerneldiag2026}.

%% file: figures/example_1_logging.tex
\begin{figure}[ht]
\centering
\begin{minipage}{0.98\columnwidth}
\begin{fullfilecode}{(a) Call Site of invalfc: kernel/cgroup/cgroup-v1.c}
return invalfc(fc, "Unknown subsys name '
\end{fullfilecode}
\vspace{-2.5mm}
\begin{fullfilecode}{(b) Declarations from logfc to invalfc: include/linux/fs\_context.h}
void logfc(struct fc_log *log, const char *prefix, char level, const char *fmt, ...);
#define __plog(p, l, fmt, ...) logfc((p)->log, (p)->prefix, l, fmt, ## __VA_ARGS__)
#define errorfc(fc, fmt, ...) __plog((&(fc)->log), 'e', fmt, ## __VA_ARGS__)
#define invalfc(fc, fmt, ...) (errorfc(fc, fmt, ## __VA_ARGS__), -EINVAL)
\end{fullfilecode}
\vspace{-2.5mm}
\begin{fullfilecode}{(c) Definition of logfc: include/linux/fs\_context.c}
void logfc(struct fc_log *log, const char *prefix, char level, const char *fmt, ...) { 
    ...
    if (!log) {
        switch (level) {
        case 'w': 
            ...
        case 'e':
            printk(KERN_ERR "
            break; 
    ...
}
\end{fullfilecode}
\end{minipage}
\caption{An example of a kernel logging macro hidden behind multi-layer macro expansion}
\label{fig:example1}
\vspace{-1.5em}
\end{figure}

%% file: scripts/4_evaluation_v2.tex
\section{Study Design}
We address the following three research questions (RQs):

\begin{itemize}[leftmargin=0.4cm]
\item \textbf{RQ1}: How accurately can \toolname{} localize the root cause?

\item \textbf{RQ2}: How effectively can \toolname{} support kernel root-cause diagnosis?
\begin{itemize}[leftmargin=0.4cm]
    \item \textbf{RQ2.1}: How effectively can KernelDiag generate high-quality root-cause explanations?
    \item \textbf{RQ2.2}: How is explanation quality related to localization correctness?
\end{itemize}

\item \textbf{RQ3}: How do the core components of KernelDiag contribute to its diagnosis performance?

\end{itemize}

\input{tables/dataset}

\subsection{Datasets}
We evaluate \toolname{} on \textit{KGYM}~\cite{mathai2024kgym}, a high-quality benchmark for Linux kernel crash diagnosis. 
It comprises 279 real-world kernel crashes spanning versions 4.x to 6.x and covering 72 distinct subsystems, including diverse logic errors across the kernel stack. 
Each case provides a complete set of diagnostic artifacts, including crash reports, reproduction scripts, execution traces, and golden patches, enabling comprehensive and reproducible evaluation.

To ensure a rigorous assessment of reasoning capability, we distinguish between cases with and without explicit root-cause hints. 
In some instances, the crash report directly reveals the faulty file or method, which simplifies the task by enabling surface-level retrieval rather than requiring deep causal reasoning. 
We therefore define two settings, \textbf{Hint} and \textbf{NoHint}, based on whether the faulty location is explicitly mentioned in the diagnostic artifacts. 
Considering both file- and method-level root-cause localization, we obtain four subsets: \textbf{F-Hint}, \textbf{F-NoHint}, \textbf{M-Hint}, and \textbf{M-NoHint}. 
These subsets are determined by regex-matching normalized ground-truth file paths or method names against the artifacts, and verified through manual inspection.
The NoHint setting represents a more realistic and challenging scenario, where the root cause must be inferred purely through cross-artifact reasoning.
This partitioning isolates superficial signal leakage from genuine reasoning, enabling a more faithful evaluation of \toolname{}'s effectiveness.
Table~\ref{tab:dataset_stats} summarizes the detailed statistics.

\subsection{Baselines}
\label{sec:baselines}

To the best of our knowledge, \toolname{} is the first approach targeting kernel-level root-cause diagnosis. 
Existing RCA approaches are primarily designed for user-space or distributed systems and are not directly applicable to the Linux kernel, due to its low-level semantics, complex control flows, and limited diagnostic signals. 
As no kernel-specific diagnosis baseline exists, we evaluate the two key aspects of root-cause diagnosis separately.
For root-cause localization, we compare \toolname{} with existing fault localization approaches; for root-cause explanation, we construct diagnosis-oriented baselines from these localization approaches.

Among existing fault localization techniques, traditional approaches are not well suited to our kernel setting. 
Spectrum-based fault localization (SBFL)~\cite{zou2019empirical, wen2019historical} often lacks the required coverage from passing and failing runs; 
mutation-based fault localization (MBFL)~\cite{jia2010analysis, andrews2006using} incurs prohibitive mutation, recompilation, and re-execution costs for such a large and configuration-sensitive system; 
and information retrieval-based fault localization (IRFL)~\cite{xia2023information, zhang2019finelocator, chen2021pathidea} struggles with sparse low-level crash reports whose symptoms are only indirectly related to the root cause. 
Besides, recent LLM-based localization approaches provide more competitive alternatives, but some of them are tightly coupled to specific programming languages or software ecosystems. 
For example, LocAgent~\cite{chen2025locagent} is a recent graph-guided LLM-agent framework for code localization, but its graph construction and indexing are tailored to Python repositories.
Adapting it to Linux kernel crash localization would require substantial re-engineering to support C language features and kernel-specific constructs, such as macros, conditional compilation, function pointers, and subsystem-specific dispatch logic, thereby introducing additional implementation choices that could confound a fair comparison.
Therefore, to assess the effectiveness of \toolname{}, we compare it against two state-of-the-art end-to-end fault localization approaches that are applicable to our kernel setting and represent the closest and most competitive alternatives:

\begin{itemize}[leftmargin=0.4cm]
    \item \linuxfl{}~\cite{zhou2025benchmarking} is the most recent and, to our knowledge, the only framework specifically designed for the Linux kernel. 
    Starting from crash reports, it extends SWE-agent~\cite{yang2024swe} with kernel-aware heuristics, including directory-aware exploration and candidate expansion,
    to better handle the scale and structural complexity of kernel codebases.
    \item \agentless{}~\cite{xia2024agentless} is a state-of-the-art LLM-based fault localization approach for general-purpose software. 
    It takes the issue report as the primary problem description and follows a three-stage pipeline:
    narrowing the search space with project-level summaries, pruning candidates through filtering, and re-ranking results with self-consistency scoring.
    By avoiding complex agent loops, \agentless{} achieves strong performance on SWE-bench\cite{jimenez2024swebench}, and serves as a robust localization baseline.
    
\end{itemize}

To ensure a fair evaluation of root-cause explanation, we construct \textit{\linuxflDiag{}} and \textit{\agentlessDiag{}}, diagnosis-oriented variants of \linuxfl{} and \agentless{}. 
Each localization method provides a Top-1 prediction and rationale, and we standardize the diagnosis stage using the same backbone LLM, prompt, and decoding settings. 
Unlike \toolname{}, which generates explanations from structured Evidence Graphs via an Explanation Agent, the baselines rely directly on the crash report and localization rationale as input.

\subsection{Metrics}
\label{sec:metrics}

We evaluate the end-to-end root-cause diagnosis effectiveness of \toolname{} in terms of root-cause localization and explanation quality.
For \textbf{root-cause localization}, 
we employ three widely adopted metrics in fault localization~\cite{li2019deepfl, zou2019empirical}:
\begin{itemize}[leftmargin=0.4cm]

\item \textbf{\textit{Top@$k$}}: the percentage of kernel crashes where the ground-truth faulty file or method is successfully localized within the Top@$k$ ranked candidates (where $k \in \{1, 3, 5, 10\}$). A higher Top@$k$ value indicates superior precision in narrowing down the search space for developers.

\item \textbf{\textit{Mean Reciprocal Rank (MRR)}}~\cite{cohen1968weighted}: the mean reciprocal ranks of the first correctly identified faulty entity: $MRR = \frac{1}{|B|} \sum_{i=1}^{|B|} \frac{1}{rank_i}$, where $|B|$ is the number of kernel crashes and $rank_i$ is the rank of the first ground-truth entity for the $i$-th crash. 
Higher MRR indicates better ranking performance.

\item \textbf{\textit{Mean First Rank (MFR)}}~\cite{li2021fault}: the average rank of the first ground-truth faulty entity across all crashes. 
For cases where no ground-truth entity appears among the Top-10 candidates, the rank is assigned a value of 11. 
A lower MFR indicates that the approach more consistently ranks the true root cause near the top of its recommendation list.
\end{itemize}

For \textbf{root-cause explanation quality}, we evaluate the explanation produced by each approach based on its Top-1 localized method, against the corresponding official patch comment.
Since patch comments emphasize fixes rather than causal mechanisms, lexical similarity metrics such as BLEU~\cite{papineni2002bleu} and ROUGE~\cite{lin2004rouge} are unsuitable. 
Following prior work~\cite{jiang2025deep, lu2025deepcrceval, sun2025bitsai, tufano2024code}, we instead conduct LLM-assisted and human evaluations across three dimensions, each rated on a 5-point Likert scale from 1 (poor) to 5 (excellent):

\begin{itemize}[leftmargin=0.4cm]
\item \textbf{\textit{Consistency}} measures whether the identified root cause aligns with the actual patch.
\item \textbf{\textit{Usefulness}} evaluates whether the root-cause explanation provides actionable guidance for fixing the root cause.
\item \textbf{\textit{Clarification}} measures whether the explanation clearly and coherently explains how the root cause led to the final crash.
\end{itemize}

\subsection{Implementation}
\toolname{} is implemented in Python using SWE-agent~\cite{yang2024swe} as the agent backbone.
For the compared localization approaches, we implement \linuxfl{} and \agentless{} following their public implementations, and evaluate them end-to-end under their original input assumptions, without manually injecting additional artifacts such as syscalls or runtime logs. 
This design preserves their original workflows and 
avoids introducing unsupported evidence integration or prompting strategies.
For the diagnosis-oriented baselines \textit{\linuxflDiag{}} and \textit{\agentlessDiag{}}, we use the same backbone LLM, temperature, diagnosis prompt, and evaluation rubric as \toolname{}, while replacing \toolname{}'s Evidence Graphs and agent rationales with the raw crash report and the localization rationales produced by \linuxfl{} or \agentless{}.

All experiments are conducted in identical virtual environments on servers equipped with 64 CPU cores and 504\, GB of memory. 
Unless otherwise specified, we use DeepSeek-V3~\cite{liu2024deepseek} as the backbone LLM with temperature 0 for fairness and reproducibility~\cite{tian2024large}.
Following prior work~\cite{zou2019empirical}, we set the candidate number $K$ to 10, aligning with the maximum workload typically acceptable to developers, and apply this setting to all baselines.

%% file: tables/dataset.tex
\begin{table}[ht]
\centering
\caption{Statistical summary of the studied KGYM dataset}
\label{tab:dataset_stats}
\begin{tabular}{@{}c|cc|c@{}}
\toprule
\textbf{Dataset} & \textbf{Hint} & \textbf{NoHint}  & \textbf{Total Crashes}\\ \midrule
File-level        & 244                            & 35                              & 279            \\
Method-level      & 164                            & 115                             & 279            \\ \bottomrule
\end{tabular}
\vspace{-0.5em}
\end{table}

%% file: scripts/5_rq1_v3.tex
\section{Results}
\subsection{RQ1: Root-Cause Localization}
\label{sec:RQ1}

\noindent\textbf{Analysis.}
To answer RQ1, we evaluate \toolname{} on the \textit{KGYM} dataset for both \emph{file-} and \emph{method-level} root-cause localization.
We compare \toolname{} with two latest baselines, \linuxfl{} and \agentless{}, spanning \textit{Hint} and \textit{NoHint} instances.
For both localization levels, we adopt standard ranking-based metrics, including Top@$k$ ($k \in \{1,3,5,10\}$), \emph{MRR}, and \emph{MFR} (detailed in Section~\ref{sec:metrics}), to evaluate the effectiveness of each approach ranking the faulty file or method.
To further demonstrate statistical significance, we perform the \textit{Friedman test}\cite{dem2006statistical} and \textit{Wilcoxon signed-rank tests}\cite{wilcoxon1945individual}.

\smallskip
\noindent\textbf{Results.}
Tables~\ref{tab:file_localization} and~\ref{tab:method_localization} report the localization performance of \toolname{}, \linuxfl{}, and \agentless{} on the \textit{All}, \textit{Hint}, and \textit{NoHint} settings at the file and method levels. 

\input{tables/1-1_file_localization}
\textbf{(1) File Level Root-cause Localization.} 
As shown in Table~\ref{tab:file_localization}, \toolname{} consistently achieves the best localization performance across all metrics, demonstrating both high accuracy and stable ranking quality. 
On the full dataset (\datafile{}), \toolname{} improves Top@1 by 27.78\% and 31.43\% over \linuxfl{} and \agentless{}, respectively, with consistent gains across Top@3–Top@10 (9.72\%–21.76\%). 
It also achieves the highest MRR (0.78) and lowest MFR (1.93), indicating that developers need to inspect fewer than two files on average, thereby reducing debugging effort in practice.

To better understand where this performance gap comes from, we further analyze the results on the \datafilenohint{} and \datafilehint{} subsets.
The results show that the weaker performance of the baselines on the full dataset is largely caused by their limited effectiveness in the challenging \datafilenohint{} setting, where no explicit file-level hints are available in the crash report.
In this subset, \linuxfl{} completely fails at Top@1 (0.00\%), while \toolname{} correctly localizes 31.43\% of faulty files.
Moreover, \toolname{} achieves substantial improvements over \linuxfl{} across Top@3--Top@10, ranging from 314.29\% to 900.00\%, and outperforms \agentless{} by 314.29\% to 1,150.00\% on the same metrics.
These results indicate the strength of \toolname{} in reasoning over incomplete and implicit evidence, rather than relying on surface-level signals.
On the easier \datafilehint{} subset, where explicit file-level hints are available, \toolname{} further pushes Top@1 accuracy to 70.90\% and outperforms \linuxfl{} and \agentless{} by 20.14\% and 23.57\%, respectively. 
Notably, \toolname{} reaches 99.18\% at Top@10, suggesting that file-level localization under hint-rich conditions is close to saturation.
Statistical analysis confirms the robustness of these improvements. 
The Friedman test indicates a significant difference among approaches ($\chi^2 = 82.819, p < 0.001$), and post-hoc Wilcoxon tests show that \toolname{} significantly outperforms both \linuxfl{} and \agentless{} ($p < 0.001$).

\input{tables/1-2_method_localization}

\textbf{(2) Method Level Root-cause Localization.} 
As shown in Table~\ref{tab:method_localization}, \toolname{} always outperforms both baselines across all datasets and evaluation metrics, demonstrating strong effectiveness at finer granularity.
\toolname{} achieves substantial improvements across all ranking metrics on the full dataset (\textit{M-All}), with Top@1–Top@10 gains of 24.09\% to 28.57\% over \linuxfl{} and 19.23\% to 34.21\% over \agentless{}.
It also delivers the best ranking quality (MRR: 0.45, MFR: 5.08), indicating that the true root cause typically appears within the top five candidates.

A more revealing pattern emerges under the \textit{M-NoHint} setting, where explicit method-level signals are absent and localization must rely purely on cross-artifact reasoning. 
In this scenario, both \linuxfl{} and \agentless{} fail to identify any correct methods at Top@1 (0.00\%), and achieve only 12.17\% and 17.39\% at Top@10, respectively. 
In contrast, \toolname{} successfully localizes 34.78\% of faulty methods within Top@10. Compared to \linuxfl{}, this corresponds to improvements ranging from 127.27\% to 185.71\% across Top@3–Top@10, and from 80.00\% to 127.27\% over \agentless{}. 
This significant gap indicates that \toolname{} can recover missing causal links and infer latent dependencies that are not explicitly reflected in the observed artifacts.
Importantly, this advantage is even more important at the method level, where the larger search space and lower signal-to-noise ratio make shallow or proximity-based heuristics less effective.
On the \textit{M-Hint} subset, \toolname{} further demonstrates strong reliability and ranking stability, achieving 53.05\% Top@1 and 95.12\% Top@10. 
This suggests that \toolname{} can not only leverage explicit hints when available, but also maintain robust causal reasoning to consistently rank the correct methods near the top.
\toolname{} significantly outperforms \linuxfl{} and \agentless{} at method-level localization (Friedman $\chi^2 = 39.795, p < 0.001$; Wilcoxon $p < 0.001$), demonstrating its consistent performance gains at method-level localization.

\finding{RQ1}{
For root-cause localization, \toolname{} consistently outperforms baselines at both file and method levels across the studied metrics (Top@k, MRR, and MFR).
In particular, at the method level, \toolname{} achieves up to 34.21\% improvement in Top-k on the full dataset and reaches 34.78\% Top@10 in the challenging NoHint setting, nearly doubling the performance of existing approaches.
}

\input{scripts/5_rq2_v3}

\input{scripts/5_rq3_v6}

%% file: tables/1-1_file_localization.tex
\begin{table}[h!]
\centering
\footnotesize
\caption{File-level Root-cause Localization Results}
\label{tab:file_localization}
\resizebox{0.99\columnwidth}{!}{
\begin{tabular}{@{}c|c|rrrr|rr}
\toprule
\multicolumn{1}{c|}{\textbf{Dataset}} &
  \multicolumn{1}{c|}{\textbf{Approach}} &
  \multicolumn{1}{c}{\textbf{Top@1}} &
  \multicolumn{1}{c}{\textbf{Top@3}} &
  \multicolumn{1}{c}{\textbf{Top@5}} &
  \multicolumn{1}{c|}{\textbf{Top@10}} &
  \multicolumn{1}{c}{\textbf{MRR}} &
  \multicolumn{1}{c}{\textbf{MFR}} \\ 
\midrule
\multirow{3}{*}{\textbf{\datafile{}}}
 &
  \textbf{\linuxflnormal{}} &
  51.61\% &
  78.85\% &
  84.95\% &
  88.53\% &
  0.66 &
  2.92 \\
 &
  \textbf{\agentlessnormal{}}
   & 50.18\%
   & 74.19\% 
   & 77.42\% 
   & 81.72\% 
   & 0.62
   & 3.50
\\
 &
  \textbf{\toolname{}} &
  \textbf{65.95\%} &
  \textbf{90.32\%} &
  \textbf{94.27\%} &
  \textbf{97.13\%} &
  \textbf{0.78} &
  \textbf{1.93} \\
\midrule
\multirow{3}{*}{\textbf{\datafilenohint{}}} 
 &
  \textbf{\linuxflnormal{}} &
  0.00\% &
  5.71\% &
  8.57\% &
  20.00\% &
  0.05 &
  9.83 \\
 &
  \textbf{\agentlessnormal{}} 
   & 0.00\%
   & 5.71\%
   & 5.71\%
   & 20.00\%
   & 0.04
   & 10.00 
   \\
 &
  \textbf{\toolname{}} &
  \textbf{31.43\%} &
  \textbf{57.14\%} &
  \textbf{71.43\%} &
  \textbf{82.86\%} &
  \textbf{0.47} &
  \textbf{4.40} \\
\midrule
\multirow{3}{*}{\textbf{\datafilehint{}}} 
 &
  \textbf{\linuxflnormal{}} &
  59.02\% &
  89.34\% &
  95.90\% &
  98.36\% &
  0.75 &
  1.93 \\
 &
  \textbf{\agentlessnormal{}} 
   & 57.38\%
   & 84.02\%
   & 87.70\%
   & 90.57\%
   & 0.71
   & 2.57
\\ 
&
  \textbf{\toolname{}} &
  \textbf{70.90\%} &
  \textbf{95.08\%} &
  \textbf{97.54\%} &
  \textbf{99.18\%} &
  \textbf{0.83} &
  \textbf{1.57} \\
  \bottomrule
\end{tabular}
}
\end{table}

%% file: tables/1-2_method_localization.tex
\begin{table}[h]
\centering
\label{tab:method_localization}
\footnotesize
\caption{Method-level Root-cause Localization Results}
\label{tab:method_localization}
\resizebox{0.99\columnwidth}{!}{
\begin{tabular}{@{}c|c|rrrr|rr}
\toprule
\multicolumn{1}{c|}{\textbf{Dataset}} &
  \multicolumn{1}{c|}{\textbf{Approach}} &
  \multicolumn{1}{c}{\textbf{Top@1}} &
  \multicolumn{1}{c}{\textbf{Top@3}} &
  \multicolumn{1}{c}{\textbf{Top@5}} &
  \multicolumn{1}{c|}{\textbf{Top@10}} &
  \multicolumn{1}{c}{\textbf{MRR}} &
  \multicolumn{1}{c}{\textbf{MFR}} \\ 
\midrule
\multirow{3}{*}{\textbf{\datamethod{}}}
 &
  \textbf{\linuxflnormal{}} &
  26.52\% &
  42.65\% &
  49.10\% &
  54.84\% &
  0.36  &
  6.30 \\
 &
  \textbf{\agentlessnormal{}} &
  27.96\% &
  40.86\% &
  47.31\% &
  57.71\% &
  0.36 &
  6.36 
\\
 &
  \textbf{\toolname{}} &
  \textbf{33.33\%} &
  \textbf{54.84\%} &
  \textbf{60.93\%} &
  \textbf{70.25\%} &
  \textbf{0.45} &
  \textbf{5.08} \\

\midrule
\multirow{3}{*}{\textbf{\datamethodnohint{}}} 
 &
  \textbf{\linuxflnormal{}} &
  0.00\% &
  6.09\% &
  9.57\% &
  12.17\% &
  0.04 &
  10.16 \\
 &
  \textbf{\agentlessnormal{}} &
  0.00\% &
  8.70\% &
  9.57\% &
  17.39\% &
  0.05 &
  9.97 
   \\
 &
  \textbf{\toolname{}} &
  \textbf{5.22\%} &
  \textbf{15.65\%} &
  \textbf{21.74\%} &
  \textbf{34.78\%} &
  \textbf{0.13} &
  \textbf{8.74} \\
\midrule
\multirow{3}{*}{\textbf{\datamethodhint{}}} 
 &
  \textbf{\linuxflnormal{}} &
  45.12\% &
  68.29\% &
  76.83\% &
  84.76\% &
  0.58 &
  3.60 \\
 &
  \textbf{\agentlessnormal{}} &
  47.56\% &
  63.41\% &
  73.78\% &
  85.98\% &
  0.58 &
  3.84 
\\ 
&
  \textbf{\toolname{}} &
  \textbf{53.05\%} &
  \textbf{82.32\%} &
  \textbf{88.41\%} &
  \textbf{95.12\%} &
  \textbf{0.68} &
  \textbf{2.51} \\
  \bottomrule
\end{tabular}
}
\end{table}

%% file: scripts/5_rq2_v3.tex
\subsection{RQ2: Root-Cause Explanation}

\medskip
\noindent
\textbf{RQ2.1: Explanation Quality}\par
\smallskip
\noindent
\textbf{Analysis.}
To evaluate the root-cause explanation quality of \toolname{}, we first conduct an LLM-assisted evaluation against the diagnostic baselines, \textit{\linuxflDiag{}} and \textit{\agentlessDiag{}}, described in Section~\ref{sec:baselines}. 
We then perform a complementary human evaluation to independently corroborate the comparative results and assess the reliability of the LLM-assisted evaluation. 
Both evaluations use an ordinal five-point Likert scale along three dimensions: \textit{Consistency}, \textit{Usefulness}, and \textit{Clarification}.
Notably, in this setting, each approach generates a textual root-cause explanation based on its Top-1 localized method, so the scores reflect the quality of the resulting end-to-end diagnosis pipeline rather than explanation quality conditioned on the correct localization.

\begin{itemize}[leftmargin=0.4cm]
    \item \textbf{LLM-assisted evaluation}: 
We adopt an \textbf{LLM-as-a-Judge} protocol using GPT-5.2, following established practices in software engineering research~\cite{jiang2025deep, lu2025deepcrceval, li2025issue, sun2025bitsai}.
For each case, the evaluator receives the official patch comment, generated explanation, ground-truth faulty method, patch diff, and evaluation rubric, and assigns a score along each of the three dimensions.
We set the temperature to 0 for reproducibility and report the average scores of all explanations generated by each approach.

    \item \textbf{Human evaluation}: 
To independently corroborate the effectiveness of \toolname{} and the LLM-assisted evaluation results, we recruit three non-author evaluators, each with over three years of experience in kernel testing.
Before annotation, we conduct a briefing session to align their understanding of the task, rubric, and annotation criteria. 
Among the 279 kernel crashes, 60 cases ($\approx 20\%$) were independently annotated by all three participants to quantify inter-rater agreement using \textit{Krippendorff's Alpha}~\cite{krippendorff2018content}, which is widely used for ordinal 5-point Likert-scale ratings~\cite{parfenova2024automating,hartvigsen2022toxigen}. 
The remaining 219 cases are evenly distributed among three evaluators. 
The resulting alpha values for \textit{Consistency}, \textit{Usefulness}, and \textit{Clarification} are 0.7506, 0.8350, and 0.7602, respectively, indicating satisfactory agreement for subjective qualitative judgments~\cite{krippendorff2018content}. 
To further examine explanation quality under different localization outcomes, we define \textbf{Hit@$k$}, where $k \in \{1,3,5,10\}$, as the subset of crashes for which the ground-truth faulty method appears in \toolname{}'s Top-$k$ method-level localization results. 
For each Hit@$k$ subset, we report the average human-assigned Likert scores of the explanations generated by \toolname{}.
\end{itemize}

\smallskip
\noindent\textbf{Results.}
Tables~\ref{tab:diag_quality_all} and~\ref{tab:rq3_score} present the diagnosis quality of \toolname{} evaluated by both GPT-5.2 and human experts across three dimensions on the \textit{KGYM} dataset.

\input{tables/3_diagnose_baselines}

\textbf{(1) LLM-assisted evaluation.}
Table~\ref{tab:diag_quality_all} shows that \toolname{} achieves higher diagnosis quality than both baselines in terms of \textit{Consistency}, \textit{Usefulness}, and \textit{Clarification}.
Compared with \linuxflDiag{}, \toolname{} improves the scores for \textit{Consistency}, \textit{Usefulness}, and \textit{Clarification} by 11.57\%, 27.97\%, and 5.76\%, respectively. 
Compared with \agentlessDiag{}, the corresponding improvements are 22.81\%, 20.90\%, and 14.40\%. 
These results indicate that \toolname{} produces diagnoses that better align with the actual root causes, provide more actionable debugging guidance, and explain fault propagation more coherently.

\textbf{(2) Human evaluation.}
Table~\ref{tab:rq3_score} shows consistent trends between human and GPT-5.2 evaluations across Hit@$k$ subsets, with \textit{Consistency}, \textit{Usefulness}, and \textit{Clarification} decreasing as $k$ increases. 
Human evaluators assign particularly high scores to \textit{Usefulness} and \textit{Consistency} at Hit@1, with average scores of 4.9677 and 4.8065, indicating that diagnoses based on correctly Top-1-localized methods closely capture the actual root causes and provide actionable guidance.
GPT-5.2 assigns lower absolute scores, particularly for \textit{Consistency} and \textit{Usefulness}, but follows the same overall trend as the human evaluation.
This difference may reflect the LLM judge's stronger reliance on official patch comments and direct repair locations and statement-level fixes, whereas humans may better recognize underlying causal mechanisms and propagation chains, even when the description differs from the patch comments.
Overall, the consistent trends across human and LLM ratings corroborate the diagnostic quality of \toolname{} and support the reliability of the LLM-assisted evaluation.

\input{tables/3_human_LLM}

\input{figures/case_study}

\textbf{(3) Root-cause explanation example.}
To further examine the quality of explanation generated by \toolname{}, we qualitatively inspect the EXT4 crash case presented in Section~\ref{ref:motivation}. 
Figure~\ref{fig:case-study} illustrates the ground-truth patch comment (a) and generated root cause explanations by \toolname{} (b).
For this case, rather than stopping at the reported crash symptom in the crash-site method  \textit{ext4\_es\_cache\_extent}, \toolname{} attributes the fault to insufficient metadata state validation in \textit{\_\_ext4\_fill\_super} during mounting, 
and correctly identifies \textit{ext4\_es\_cache\_extent} as only the downstream crash site where the violated extent invariant (\texttt{end < lblk}) is exposed.
Importantly, instead of merely restating the \texttt{BUG()} location,
it reconstructs a propagation chain from corrupted descriptors/metadata at mount, to lookup/traversal operations, and finally the extent caching crash.
This explanation is valuable because, unlike the crash report alone, which reveals only the terminal execution state, it provides a substantially more actionable debugging hypothesis by linking the earlier anomaly, the intermediate invalid state, and the final failure into a coherent causal narrative.

\finding{RQ2.1}{
Both LLM-assisted and human evaluations show consistent trends, demonstrating that \toolname{} maintains stable root-cause explanation quality and consistently generates reliable, informative, and interpretable explanations.
}

\medskip
\noindent
\textbf{RQ2.2: Localization--Explanation Relationship}\par

\smallskip
\noindent
\textbf{Analysis.}
To examine how localization correctness relates to explanation quality, we divide the cases of each approach by whether the Top-1 predicted method matches the ground truth and compare their average \textit{Consistency}, \textit{Usefulness}, and \textit{Clarification} scores evaluated by GPT-5.2.
We further compare score distributions using two-sided \textit{Mann--Whitney U tests}~\cite{mann1947test}, apply \textit{Holm correction} for multiple comparisons~\cite{holm1979simple}, and report \textit{Cliff's} $\delta$ as the effect size~\cite{cliff1993dominance}.

\smallskip
\noindent
\textbf{Results.}
Table~\ref{tab:rq2_relationship_v1} reports case counts and average \textit{Consistency}, \textit{Usefulness}, and \textit{Clarification} scores for \linuxflDiag{}, \agentlessDiag{}, and \toolname{} under correct (\checkmark) and incorrect ($\times$) Top-1 localization on the \textit{KGYM} dataset.
Across all approaches, correct Top-1 localization is associated with higher \textit{Consistency} and \textit{Usefulness}.
For \textit{Consistency}, correct cases exceed incorrect ones by 1.2494, 0.8238, and 1.1613 score points for \linuxflDiag{}, \agentlessDiag{}, and \toolname{}, respectively. 
The \textit{Mann--Whitney U tests} confirm that all differences remain significant after Holm correction (adjusted $p<0.001$), with \textit{large}, \textit{small}, and \textit{medium} effects, respectively: \linuxflDiag{} ($U=11{,}360.5$, $\delta=0.4978$), \agentlessDiag{} ($U=10{,}410.0$, $\delta=0.3280$), and \toolname{} ($U=12{,}735.0$, $\delta=0.4724$).
For \textit{Usefulness}, 
correct cases exceed incorrect ones 
by 1.3472, 0.9384, and 1.1935 score points for \linuxflDiag{}, \agentlessDiag{}, and \toolname{}, respectively. 
These differences are also statistically significant after Holm correction (all adjusted $p<0.001$), with \textit{medium} effects for \linuxflDiag{} ($U=11{,}078.0$, $\delta=0.4605$) and \agentlessDiag{} ($U=10{,}845.5$, $\delta=0.3835$), and a \textit{large} effect for \toolname{} ($U=13{,}170.0$, $\delta=0.5227$). 
These results provide statistical evidence that correct root-cause localization is associated with explanations that better align with the actual fault and provide more useful debugging information.

\input{tables/3_relationship_v1}

However, \textit{Clarification} shows no significant difference
between correct and incorrect Top-1 cases 
after Holm correction, with adjusted $p$-values of 0.595, 0.595, and 0.409 for \linuxflDiag{}, \agentlessDiag{}, and \toolname{}, respectively. 
The corresponding effect sizes are all negligible ($\delta=-0.0222$, 0.0408, and 0.0845). 
One possible reason is that an LLM can construct a coherent fault-propagation narrative even when it is centered on an incorrect method; therefore, \textit{Clarification} alone does not indicate that an explanation is grounded in the actual root cause.

\finding{RQ2.2}{
Correct Top-1 root-cause localization is significantly associated with higher \textit{Consistency} and \textit{Usefulness} across all approaches, with \textit{small}-to-\textit{large} effects, while its association with \textit{Clarification} is negligible.
Accurate root-cause localization therefore provides an important foundation for explanations that align with the actual fault and support debugging.
}

%% file: tables/3_diagnose_baselines.tex
\begin{table}[t]
\centering
\caption{Root-cause Explanation Quality Scores Assessed by GPT-5.2}
\label{tab:diag_quality_all}
\begin{tabular}{c|ccc}
\toprule
  \multicolumn{1}{c|}{\textbf{Approach}} &
  \multicolumn{1}{c}{\textbf{Consistency}} &
  \multicolumn{1}{c}{\textbf{Usefulness}} &
  \multicolumn{1}{c}{\textbf{Clarification}} \\
  \midrule

{\textbf{\linuxflDiag}} &
  {2.6631} &
  {2.2939} &
  {3.9857}
\\ 
{\textbf{\agentlessDiag}} &
  {2.4194} &
  {2.4265} &
  {3.6846}
\\ 
\textbf{\toolname{}} &
  \textbf{2.9713} &
  \textbf{2.9355} &
  \textbf{4.2151} \\
\bottomrule
\end{tabular}
\vspace{-1em}
\end{table}

%% file: tables/3_human_LLM.tex
\begin{table}[t]
\centering
\caption{Human \& LLM Evaluation of \toolname{}'s Explanations}
\label{tab:rq3_score}
\resizebox{0.99\columnwidth}{!}{
\begin{tabular}{c|c|
  @{\hspace{5pt}}c
  *{4}{@{\hspace{5pt}}c}
  }
\toprule
\textbf{Dimension} &
  \textbf{Evaluator} &
  \multicolumn{1}{c}{\textbf{\textbf{Hit@1}}} &
  \multicolumn{1}{c}{\textbf{\textbf{Hit@3}}} &
  \multicolumn{1}{c}{\textbf{\textbf{Hit@5}}} &
  \multicolumn{1}{c}{\textbf{\textbf{Hit@10}}}\\ 
  
  \midrule
\multirow{2}{*}{\textbf{Consistency}}   
& Human & 4.8065 & 3.8889 & 3.6647 & 3.3265  \\
& GPT-5.2   & 3.7312 & 3.5033 & 3.4000 & 3.2398  \\
\midrule

\multirow{2}{*}{\textbf{Usefulness}}    
& Human & 4.9677 & 4.5098 & 4.3412 & 4.1071  \\ 
& GPT-5.2    & 3.7312 & 3.4837 & 3.3647 & 3.2194 \\
\midrule
\multirow{2}{*}{\textbf{Clarification}} 
& Human & 4.5484 & 4.3464 & 4.3000 & 4.2194 \\ 
& GPT-5.2 &   4.2796 & 4.3072& 4.3059& 4.2806\\

\bottomrule
\end{tabular}
}
\vspace{-0.5em}
\end{table}

%% file: figures/case_study.tex
\begin{figure}[h]
\centering
\begin{minipage}{0.98\columnwidth}
\begin{textbox}{(a) Ground-truth from the Patch Comment}
...
The problem with using EXT4\_FC\_REPLAY is that a maliciously corrupted
superblock could result in EXT4\_FC\_REPLAY getting set in
s\_mount\_state. This bypasses some sanity checks, and this can trigger
a BUG() in ext4\_es\_cache\_extent(). 
\end{textbox}
\vspace{-2.5mm}
\begin{textbox}{(b) Generated Root-cause Explanation by \toolname{}}
The issue occurs because \_\_ext4\_fill\_super fails to properly validate metadata state during mount when group descriptor checksums are corrupted. Even though ext4\_check\_descriptors detects the corruption, mounting continues without journal recovery, allowing invalid extent states to propagate. 

The Propagation Chain to the Crash is: 

1. Corrupted group descriptors detected during mount 

2. Invalid extent metadata not caught by mount-time validation 

3. Directory lookup triggers extent tree traversal

4. ext4\_es\_cache\_extent encounters invalid extent (end $<$ lblk) 

5. BUG\_ON triggers due to violated invariant
\end{textbox}
\end{minipage}
\caption{The example root-cause explanation generated by \toolname{}}
\label{fig:case-study}
\vspace{-1em}
\end{figure}

%% file: tables/3_relationship_v1.tex
\begin{table}[t]
\centering
\caption{Explanation Quality under Correct and Incorrect Top-1 Root-Cause Localization}
\label{tab:rq2_relationship_v1}
\resizebox{0.99\columnwidth}{!}{
\begin{tabular}{c|c|c|ccc}
\toprule
\multirow{2}{*}{\textbf{Approach}} & \textbf{Top-1} & \multirow{2}{*}{\textbf{\#Cases}} & \multirow{2}{*}{\textbf{Consistency}} & \multirow{2}{*}{\textbf{Usefulness}} & \multirow{2}{*}{\textbf{Clarification}} \\
& \textbf{Correctness} & & & & \\
\midrule
\multirow{2}{*}{\textbf{\linuxflDiag{}}}  & \checkmark   & 74  & 3.5811 & 3.2838 & 3.9595 \\
                                & $\times$ & 205 & 2.3317 & 1.9366 & 3.9951 \\
                                \midrule
\multirow{2}{*}{\textbf{\agentlessDiag{}}} & \checkmark   & 78  & 3.0128 & 3.1026 & 3.7308 \\
                                & $\times$ & 201 & 2.1891 & 2.1642 & 3.6667 \\
                                \midrule
\multirow{2}{*}{\textbf{\toolname{}}}     & \checkmark   & 93  & 3.6452 & 3.7312 & 4.2796 \\
                                & $\times$ & 186 & 2.4839 & 2.5376 & 4.1828
                                \\
\bottomrule
\end{tabular}
}
\vspace{-1em}
\end{table}

%% file: scripts/5_rq3_v6.tex
\subsection{RQ3: Component Contribution}
\label{sec:ablation}
\noindent\textbf{Analysis.}
To answer RQ3, we construct controlled variants of \toolname{} to assess the contributions of (i) heterogeneous artifacts, (ii) the reasoning framework, and (iii) key architectural components.
Each variant removes or replaces one design element to isolate its performance impact:

\begin{itemize}[leftmargin=0.4cm]
    \item \textbf{Heterogeneous Artifact}: We construct three variants (\textbf{\toolwosyscall}, \textbf{\toolwolog}, and \textbf{\toolworeport}), each removing one artifact type (syscalls, logs, or crash reports, respectively). 
    This setting evaluates each artifact's contribution to causal-chain reconstruction and whether removing any single source causes incomplete or misleading reasoning.
    \item \textbf{Reasoning Framework}: 
    We examine two simplified reasoning strategies.
    \textbf{\toolnoagent} bypasses structured reasoning entirely by directly querying the LLM in a single-pass manner, without repository-level exploration or intermediate evidence construction by agents.
    \textbf{\toolsingleagent} merges all heterogeneous inputs into a single agent (based on a vanilla SWE-agent), removing role specialization and forcing unified reasoning over mixed-format artifacts.
    These variants assess the importance of decomposition and role-based reasoning in handling heterogeneous evidence.
    \item \textbf{Architectural Component}:
    We ablate four key mechanisms.
    \textbf{\toolnomapping} disables log-to-code mapping by removing logging location and severity, preventing logs from being grounded to source-level context.
    \textbf{\toolnointrospection} disables the custom command-line tools for retrieving complete function implementations and macro definitions, retaining only basic shell utilities and SWE-agent's default repository-navigation tools, such as \texttt{find}, \texttt{grep}, \texttt{ls}, \texttt{find\_file}, and \texttt{search\_dir}.
    \textbf{\toolnoconfig} withholds the environment artifacts such as kernel \texttt{.config}, forcing agents to infer configuration-dependent paths from source context and prior knowledge without verifying their feasibility in the target environment.
    \textbf{\toolnograph} replaces the structured evidence graph with raw textual outputs from individual agents, eliminating explicit causal modeling and cross-source alignment.
    These variants evaluate whether precise grounding, in-depth exploration and structured representation are necessary for effective root-cause reasoning.

\end{itemize}

\input{tables/2_ablation_all}
\input{tables/2_ablation_all_explanation}

\smallskip
\noindent
\textbf{Results.}
Tables~\ref{tab:ablation} and~\ref{tab:ablation_explanation} report the localization and explanation results of \toolname{} and its variants.
Overall, \toolname{} is most sensitive to the removal of heterogeneous artifacts, followed by the reasoning framework and architecture components. 
Among the artifacts, crash reports contribute most substantially: removing them (\textbf{\textit{\toolworeport{}}}) reduces file-/method-level Top@1 by $63.04\%$/$78.49\%$ and \textit{Consistency} by $61.76\%$, confirming their role as the primary causal anchor. 
Syscalls and logs provide complementary evidence, with logs being particularly important on \datamethodnohint{} (Top@1: $-50.00\%$).
Within the reasoning framework, replacing agent-based reasoning with a single-pass manner (\textbf{\textit{\toolnoagent{}}}) causes the largest degradation (Top@1: $-28.26\%$/$-29.03\%$; \textit{Usefulness}: $-26.86\%$), while the performance degradation of \textbf{\textit{\toolsingleagent{}}} highlights the benefit of role specialization. 
Among the architecture-component ablations, \textbf{\textit{\toolnomapping{}}} causes the largest method-level Top@1 drop ($-12.90\%$), followed by \textbf{\textit{\toolnoconfig{}}} ($-11.83\%$), \textbf{\textit{\toolnograph{}}} ($-10.75\%$), and \textbf{\textit{\toolnointrospection{}}} ($-8.60\%$), demonstrating the importance of source grounding, feasible-path filtering, structured causal modeling, and fine-grained semantic inspection.

\finding{RQ3}{
Ablation results indicate that the root-cause diagnosis gains of \toolname{} arise from the joint contribution of its components, as removing any element consistently degrades both root-cause localization effectiveness and root-cause explanation quality.
}

%% file: tables/2_ablation_all.tex
\begin{table}[t]
\centering
\caption{Root-cause localization results of Ablation Study
}
\label{tab:ablation}
\resizebox{.99\linewidth}{!}{
\begin{tabular}{@{}c|c|rrrr|cc@{}}
\toprule
\multicolumn{1}{c|}{\textbf{Dataset}} &
  \multicolumn{1}{c|}{\textbf{Approach}} &
  \multicolumn{1}{c}{\textbf{Top@1}} &
  \multicolumn{1}{c}{\textbf{Top@3}} &
  \multicolumn{1}{c}{\textbf{Top@5}} &
  \multicolumn{1}{c|}{\textbf{Top@10}} &
  \multicolumn{1}{c}{\textbf{MRR}} &
  \multicolumn{1}{c}{\textbf{MFR}} \\  \midrule
\multirow{10}{*}{\textbf{\datafile{}}}  & \textbf{\toolname{}}        & \textbf{65.95\%} & \textbf{90.32\%} & \textbf{94.27\%} & \textbf{97.13\%} & \textbf{0.78} & \textbf{1.93} \\ 
                      & {\color{groupA} \textbf{\toolwosyscall{}}}   & {\color{groupA}59.86\%}          & {\color{groupA}82.80\%}          & {\color{groupA}85.66\%}          & {\color{groupA}88.53\%}          & {\color{groupA}0.71}          & {\color{groupA}2.72}          \\
                      & {\color{groupA}\textbf{\toolwolog{}}}       & {\color{groupA}60.57\%}          & {\color{groupA}84.95\%}          & {\color{groupA}91.40\%}          & {\color{groupA}94.27\%}          & {\color{groupA}0.73}          & {\color{groupA}2.30}          \\
                      & {\color{groupA}\textbf{\toolworeport{}}}    & {\color{groupA}24.37\%}              & {\color{groupA}43.73\%}              & {\color{groupA}53.05\%}           & {\color{groupA}59.14\%}            & {\color{groupA}0.36}              & {\color{groupA}6.09} \\

                      & {\color{groupB}\textbf{\toolnoagent{}}}    & {\color{groupB}47.31\%}          & {\color{groupB}66.31\%}          & {\color{groupB}72.40\%}          & {\color{groupB}75.99\%}          & {\color{groupB}0.58}          & {\color{groupB}4.09}          \\
                      & {\color{groupB}\textbf{\toolsingleagent{}}} & {\color{groupB}56.99\%}          & {\color{groupB}82.44\%}          & {\color{groupB}89.61\%}          & {\color{groupB}92.83\%}          & {\color{groupB}0.71}          & {\color{groupB}2.47}          \\
                      

                      &  {\color{groupC}\textbf{\toolnomapping{}}}   & {\color{groupC}59.50\%}          & {\color{groupC}87.46\%}          & {\color{groupC}91.40\%}          & {\color{groupC}94.98\%}          & {\color{groupC}0.74}          & {\color{groupC}2.23}       \\ 
                      &  {\color{groupC}\textbf{\toolnointrospection{}}}   & {\color{groupC}56.99\%}          & {\color{groupC}85.30\%}          & {\color{groupC}91.04\%}          & {\color{groupC}93.55\%}          & {\color{groupC}0.72}          & {\color{groupC}2.34}       \\ 
                      
                      &  {\color{groupC}\textbf{\toolnoconfig{}}}   & {\color{groupC}58.78\%}          & {\color{groupC}83.51\%}          & {\color{groupC}90.68\%}          & {\color{groupC}93.55\%}          & {\color{groupC}0.72}          & {\color{groupC}2.41}       \\ 
                    
                      & {\color{groupC}\textbf{\toolnograph{}}}     & {\color{groupC}53.41\%}          & {\color{groupC}80.29\%}          & {\color{groupC}85.30\%}          & {\color{groupC}89.25\%}          & {\color{groupC}0.68}          & {\color{groupC}2.83}          \\ 
                      
                      \midrule
                      \multirow{10}{*}{\textbf{\datamethod{}}}  & \textbf{\toolname{}}       & \textbf{33.33\%} & \textbf{54.84\%} & \textbf{60.93\%} & \textbf{70.25\%} & \textbf{0.45} & \textbf{5.08} \\ 
                      & {\color{groupA}\textbf{\toolwosyscall{}}}   & {\color{groupA}30.11\%}          & {\color{groupA}49.10\%}          & {\color{groupA}55.56\%}          & {\color{groupA}62.72\%}          & {\color{groupA}0.41}          & {\color{groupA}5.67}     \\
                      & {\color{groupA}\textbf{\toolwolog{}}}       & {\color{groupA}29.03\%}          & {\color{groupA}48.03\%}          & {\color{groupA}56.27\%}          & {\color{groupA}65.23\%}          & {\color{groupA}0.41}          & {\color{groupA}5.65}     \\
                      & {\color{groupA}\textbf{\toolworeport{}}}    & {\color{groupA}7.17\%}           & {\color{groupA}11.83\%}          & {\color{groupA}16.13\%}          & {\color{groupA}21.51\%}          & {\color{groupA}0.11}          & {\color{groupA}9.39}     \\ 
                      & {\color{groupB}\textbf{\toolnoagent{}}}     & {\color{groupB}23.66\%}          & {\color{groupB}38.71\%}          & {\color{groupB}43.37\%}          & {\color{groupB}48.39\%}          & {\color{groupB}0.32}          & {\color{groupB}6.85}          \\
                      & {\color{groupB}\textbf{\toolsingleagent{}}} & {\color{groupB}31.90\%}          & {\color{groupB}50.18\%}          & {\color{groupB}55.56\%}          & {\color{groupB}65.59\%}          & {\color{groupB}0.42}          & {\color{groupB}5.56}          \\
                      & {\color{groupC}\textbf{\toolnomapping{}}}   & {\color{groupC}29.03\%}          & {\color{groupC}53.41\%}          & {\color{groupC}59.14\%}          & {\color{groupC}65.59\%}          & 
                      {\color{groupC}0.42}          & 
                      {\color{groupC}5.39}          \\
                      & {\color{groupC}\textbf{\toolnointrospection{}}}   & {\color{groupC}30.47\%}          & {\color{groupC}50.18\%}          & {\color{groupC}54.84\%}          & {\color{groupC}65.23\%}          & 
                      {\color{groupC}0.41}          & 
                      {\color{groupC}5.59}          \\
                      & {\color{groupC}\textbf{\toolnoconfig{}}}   & {\color{groupC}29.39\%}          & {\color{groupC}49.10\%}          & {\color{groupC}56.27\%}          & {\color{groupC}65.59\%}          & 
                      {\color{groupC}0.41}          & 
                      {\color{groupC}5.61}          \\
                      & {\color{groupC}\textbf{\toolnograph{}}}     & {\color{groupC}29.75\%}          & {\color{groupC}47.67\%}          & {\color{groupC}54.84\%}          & {\color{groupC}63.44\%}          & 
                      {\color{groupC}0.41}          & 
                      {\color{groupC}5.68}          \\ 
                       \bottomrule
\end{tabular}
}
\vspace{-1em}
\end{table}

%% file: tables/2_ablation_all_explanation.tex
\begin{table}[t]
\centering
\caption{Root-cause explanation results of Ablation Study
}
\label{tab:ablation_explanation}
\resizebox{.99\linewidth}{!}{
\begin{tabular}{c|ccc}

\toprule
\textbf{Approach}      & \textbf{Consistency} & \textbf{Usefulness} & \textbf{Clarification} \\
\midrule
\textbf{\toolname{}}    & \textbf{2.9713}      & \textbf{2.9355}     & \textbf{4.2151}        \\
{\color{groupA}\textbf{\toolwosyscall{}}}    & {\color{groupA}2.5771}      & {\color{groupA}2.6882}     & {\color{groupA}4.0681}        \\
{\color{groupA}\textbf{\toolwolog{}}}        & {\color{groupA}2.6201}      & {\color{groupA}2.7455}     & {\color{groupA}4.0932}        \\
{\color{groupA}\textbf{\toolworeport{}}}     & {\color{groupA}1.1362}      & {\color{groupA}1.1864}     & {\color{groupA}3.9713}        \\
{\color{groupB}\textbf{\toolnoagent{}}}       & {\color{groupB}2.2939}      & {\color{groupB}2.1470}     & {\color{groupB}3.7885}        \\
{\color{groupB}\textbf{\toolsingleagent{}}}   & {\color{groupB}2.5448}      & {\color{groupB}2.7097}     & {\color{groupB}4.0000}        \\
{\color{groupC}\textbf{\toolnomapping{}}}  & {\color{groupC}2.5197}      & {\color{groupC}2.6595}     & {\color{groupC}4.0717}        \\
{\color{groupC}\textbf{\toolnointrospection{}}}  & {\color{groupC}2.7168}      & {\color{groupC}2.7706}     & {\color{groupC}4.2007}        \\ 	 	 
{\color{groupC}\textbf{\toolnoconfig{}}}  & {\color{groupC}2.6953}      & {\color{groupC}2.8100}     & {\color{groupC}4.0609}        \\		
{\color{groupC}\textbf{\toolnograph{}}}     & {\color{groupC}2.4839}      & {\color{groupC}2.6380}     & {\color{groupC}4.0932}       \\
\bottomrule
\end{tabular}
}
\vspace{-1em}
\end{table}

%% file: scripts/6_discussion.tex
\section{Discussion}
\label{sec:discussion}

\smallskip
\noindent
\textbf{I. Data Leakage Analysis. } 
Our experiments use DeepSeek-V3, released in December 2024, whereas the latest \textit{KGYM} bug was fixed by 2023~\cite{mathai2024kgym}. 
Consequently, some kernel crashes or their fixes may have appeared in the model's pre-training corpus.
To assess this threat, we further evaluate \toolname{} on 50 kernel crashes reported and fixed after March 2025, ensuring their unavailability during the model's pre-training. 
This additional post-release dataset includes challenging \textit{NoHint} cases, 
where diagnostic artifacts omit explicit references to faulty files (8\%) and faulty methods (24\%).

For root-cause localization, as shown in Table~\ref{tab:leakage_loc_res}, which reports file- and method-level localization performance on the post-release dataset, \toolname{} consistently and substantially outperforms both \linuxfl{} and \agentless{} across all localization metrics. At the file level, \toolname{} improves Top@1 accuracy by 20.69\% and 12.90\% over \linuxfl{} and \agentless{}, respectively, while maintaining a consistent advantage across Top@3–Top@10, with gains ranging from 2.04\% to 17.07\%. Notably, \toolname{} achieves 100\% Top@10, successfully placing the root-cause files within the top 10 candidates for all 50 crashes, including the four challenging \textit{NoHint} cases. At the method level, \toolname{} yields even larger improvements, outperforming \linuxfl{} and \agentless{} by 60.00\% and 26.32\% in Top@1 accuracy and maintaining gains of 19.44\%–62.50\% across Top@3–Top@10. Furthermore, \toolname{} achieves the highest MRR (0.62) and the lowest MFR (3.28), indicating that developers need to inspect only about three methods on average to identify the root cause.

\input{tables/4_leak_localization_res}
\input{tables/4_leak_diag_res}

Table~\ref{tab:leakage_diag_res} presents the corresponding root-cause explanation quality, evaluated by GPT-5.2 on three dimensions: Consistency, Usefulness, and Clarification. 
A similar trend is observed, with \toolname{} achieving the highest scores across all dimensions. 
In particular, compared with \linuxfl{} and \agentless{}, \toolname{} improves \textit{Consistency} by 13.53\%/7.86\%, \textit{Usefulness} by 23.08\%/16.79\%, and \textit{Clarification} by 11.48\%/9.68\%, respectively. 
These improvements suggest that the performance gains of \toolname{} are unlikely to arise from memorization of historical crash data. 
Instead, they indicate that \toolname{} derives its advantages from structured reasoning over heterogeneous evidence sources, enabling effective root-cause localization and explanation generation even for previously unseen crashes.

\smallskip
\noindent
\textbf{II. Generalizability on Different LLM Backends.}
To assess the sensitivity of \toolname{} to the LLM backends, we further evaluate \toolname{} with Qwen3-Max~\cite{qwen3max} on the \textit{KGYM} dataset. 
Tables~\ref{tab:qwen3_localization} and \ref{tab:qwen3_diagnosis} report the root-cause localization and explanation-quality results, respectively, using Qwen3-Max as the LLM backend.
Despite minor performance variations across different backends, \toolname{} remains highly competitive.
It surpasses existing baselines in root-cause localization and delivers better explanation quality in terms of \textit{Consistency}, \textit{Usefulness}, and \textit{Clarification}.
These results show that the gains of \toolname{} are not tied solely to a particular LLM, but are largely attributable to its kernel-specific design.

\input{tables/4_qwen3_localization}

\input{tables/4_qwen3_diagnosis}

We further examine the overlap between the cases correctly localized at Top@1 by the two LLM backends. 
At the file level, both backends correctly localize 146 common cases at Top@1, while DeepSeek-V3 and Qwen3-Max uniquely identify 38 and 31 cases, respectively.
At the method level, they overlap on 59 Top@1 cases, with 34 and 43 unique to each backend. 
These results indicate that different LLM backends capture partially complementary reasoning signals, suggesting that multi-backend diagnosis may further improve robustness in practice.

\smallskip
\noindent
\textbf{III. Cost–Accuracy Trade-offs.} 
\toolname{} tends to incur higher time and token costs due to role-specialized agent-based reasoning over heterogeneous artifacts and on-demand retrieval of fine-grained repository context. 
We view this overhead as a deliberate trade-off for more accurate and comprehensive diagnosis. 
Unlike IR-based approaches~\cite{xia2023information, zhang2019finelocator, chen2021pathidea, razzaq2021boostnsift} that rely on shallow textual matching, \toolname{} more closely follows the diagnostic workflow of human engineers by jointly interpreting heterogeneous artifacts and grounding them in detailed source code, which is essential when the failure mechanism cannot be recovered from any single artifact alone.
In practice, artifact-specific analyses can run in parallel to reduce latency, and inference costs are typically small relative to the engineering effort required for manual crash reproduction and inspection of reports, syscalls, logs, and code.
Therefore, for real-world kernel diagnosis, where accuracy and reduced mean time to repair (MTTR) are critical, 
the additional computational cost is justified.

\smallskip
\noindent
\textbf{IV. Failure Analysis.}
Figure~\ref{fig:failure_diagnosis} shows an out-of-bounds memory crash\footnote{https://syzkaller.appspot.com/bug?extid=45862e7027be5d590577} that \toolname{} fails to diagnose.
According to the patch, the ground truth is that the update of frame-buffer geometry makes the computed margin sizes valid, while the original code only checks them as nonzero values, thereby allowing an integer-underflow-induced invalid width/height to propagate further, and eventually leads to an out-of-bounds memory access in another subsystem.
In this case, the crash report shows the final observed failure site; 
the syscall shows an open of \texttt{/dev/fb0} and a frame-buffer update;
the runtime log confirms access to the device.
Based on these artifacts, \toolname{} narrows the diagnosis to a suspicious frame-buffer state update, but fails to recover the root cause
as it lacks the intermediate steps connecting the configuration update to the later invalid size calculation.
This failure diagnosis suggests that relying only on preexisting diagnostic artifacts is insufficient for such cases. 
Future work could combine our approach with lightweight dynamic analysis (e.g., breakpoints or targeted logging in suspicious functions) to capture missing execution details and reconstruct the fault propagation path.

\input{figures/failure_analysis}

\section{Threats to Validity}
\noindent
\textbf{Internal Validity.} 
The main internal threat is potential data leakage from LLM pre-training, since the bugs in \textit{KGYM} were fixed before the release of DeepSeek-V3. 
We mitigate this concern through the post-release evaluation and the LLM-backend analysis in Section~\ref{sec:discussion}, and the ablation results in Section~\ref{sec:ablation}  show that direct-query and single-agent variants perform worse than the full approach, suggesting that memorization alone is insufficient. 
Nevertheless, as the exact pre-training corpora of closed-source LLMs are not transparent, data leakage cannot be completely ruled out.

\smallskip
\noindent
\textbf{External Validity.} 
Our experiments are mainly conducted on \textit{KGYM}, which covers diverse real-world Linux kernel crashes. 
Nevertheless, \textit{KGYM} may not fully capture the variability across kernel versions, hardware architectures, and subsystem-specific behaviors encountered in production environments. 
In addition, our evaluation focuses exclusively on Linux kernel bugs, and thus the results may not directly generalize to other low-level systems.
Although the additional post-release dataset increases temporal diversity and reduces the risk of benchmark-specific conclusions, further studies on broader datasets and software ecosystems are necessary to establish the general applicability of \toolname{}.

\smallskip
\noindent
\textbf{Construct Validity.} 
The primary threat concerns the subjectivity of manual evaluation in assessing diagnosis quality.
To mitigate this, we recruited three independent evaluators with experience in Linux kernel testing, none of whom were involved in this work. 
We further quantify inter-rater reliability using \textit{Krippendorff’s Alpha}, ensuring a high level of agreement and consistency in the assessments.

%% file: tables/4_leak_localization_res.tex
\begin{table}[t!]
\centering
\caption{Root-cause Localization Results on the \textbf{Post-release Dataset}}
\label{tab:leakage_loc_res}
\resizebox{.99\columnwidth}{!}{
\begin{tabular}{@{}c|c|rrrr|cc@{}}
\toprule
\multicolumn{1}{c|}{\textbf{Dataset}} &
  \multicolumn{1}{c|}{\textbf{Approach}} &
  \multicolumn{1}{c}{\textbf{Top@1}} &
  \multicolumn{1}{c}{\textbf{Top@3}} &
  \multicolumn{1}{c}{\textbf{Top@5}} &
  \multicolumn{1}{c|}{\textbf{Top@10}} &
  \multicolumn{1}{c}{\textbf{MRR}} &
  \multicolumn{1}{c}{\textbf{MFR}} \\  \midrule
\multirow{3}{*}{\textbf{\datafile{}}}  
                       & \textbf{\linuxflnormal{}}   & 58.00\% & 92.00\% & 96.00\% & 98.00\% & 0.74  & 1.98 \\ 
                       & \textbf{\agentlessnormal{}} & 62.00\% & 82.00\% & 84.00\%  & 90.00\%  & 0.72 & 2.74 \\
                       & \textbf{\toolname{}} & \textbf{70.00\%} & \textbf{96.00}\% & \textbf{98.00\%} & \textbf{100.00\%} & \textbf{0.82} & \textbf{1.52} \\
\midrule
\multirow{3}{*}{\textbf{\datamethod{}}}  
                       & \textbf{\linuxflnormal{}}   & 30.00\% & 48.00\% &  62.00\% &  66.00\% &  0.42 & 5.38 \\
                       & \textbf{\agentlessnormal{}} & 38.00\% & 52.00\% & 66.00\%  & 72.00\%  & 0.48 & 4.80 \\
                       & \textbf{\toolname{}} & \textbf{48.00\%} & \textbf{78.00\%} & \textbf{80.00\%}  & \textbf{86.00\%}  & \textbf{0.62} & \textbf{3.28} \\
                       
\bottomrule
\end{tabular}
}
\vspace{-0.5em}
\end{table}

%% file: tables/4_leak_diag_res.tex
\begin{table}[t]
\centering
\caption{
Root-Cause Explanation Quality on the \textbf{Post-release Dataset}
}
\label{tab:leakage_diag_res}
\begin{tabular}{c|ccc}
\toprule
  \multicolumn{1}{c|}{\textbf{Approach}} &
  \multicolumn{1}{c}{\textbf{Consistency}} &
  \multicolumn{1}{c}{\textbf{Usefulness}} &
  \multicolumn{1}{c}{\textbf{Clarification}} \\
  \midrule
{\textbf{\linuxflDiag}} &
  {2.6600} &
  {2.6000} &
  {3.6600}
\\  	 	 	
{\textbf{\agentlessDiag}} &
  {2.8000} &
  {2.7400} &
  {3.7200}
\\  
\textbf{\toolname{}} &
  \textbf{3.0200} &
  \textbf{3.2000} &
  \textbf{4.0800} \\
\bottomrule
\end{tabular}
\vspace{-0.5em}
\end{table}

%% file: tables/4_qwen3_localization.tex
\begin{table}[t]
\centering
\caption{Root-cause Localization Results Using \textbf{Qwen3-Max}}
\label{tab:qwen3_localization}
\resizebox{.99\columnwidth}{!}{
\begin{tabular}{c|c|rrrr|rr}
\toprule
\textbf{Dataset} &
  \textbf{Approach} &
  \multicolumn{1}{c}{\textbf{Top@1}} &
  \multicolumn{1}{c}{\textbf{Top@3}} &
  \multicolumn{1}{c}{\textbf{Top@5}} &
  \multicolumn{1}{c|}{\textbf{Top@10}} &
  \multicolumn{1}{c}{\textbf{MRR}} &
  \multicolumn{1}{c}{\textbf{MFR}} \\ 
  \midrule
\multirow{3}{*}{\textbf{\datafile}}   & \textbf{\linuxflnormal}   & 55.91\%          & 82.44\%          & 86.74\%          & 90.32\%          & 0.69          & 2.72          \\
                                      & \textbf{\agentlessnormal} & 48.75\%          & 68.10\%          & 72.04\%          & 72.40\%          & 0.59          & 4.14          \\
                                      & \textbf{\toolname}        & \textbf{63.44\%} & \textbf{89.25\%} & \textbf{93.55\%} & \textbf{96.06\%} & \textbf{0.77} & \textbf{2.00} \\ 
                                       \midrule
\multirow{3}{*}{\textbf{\datamethod}} & \textbf{\linuxflnormal}   & 29.39\%          & 44.44\%          & 52.33\%          & 60.22\%          & 0.39          & 5.95          \\
                                      & \textbf{\agentlessnormal} & 29.75\%          & 41.22\%          & 47.31\%          & 58.06\%          & 0.38          & 6.25          \\
                                      & \textbf{\toolname}        & \textbf{36.56\%} & \textbf{59.50\%} & \textbf{64.87\%} & \textbf{75.99\%} & \textbf{0.49} & \textbf{4.62} \\
\bottomrule
                                      
\end{tabular}
}
\vspace{-0.5em}
\end{table}

%% file: tables/4_qwen3_diagnosis.tex
\begin{table}[t]
\centering
\caption{Average Explanation Quality Scores Using \textbf{Qwen3-Max}}
\label{tab:qwen3_diagnosis}
\begin{tabular}{c|ccc}
\toprule
  \multicolumn{1}{c|}{\textbf{Approach}} &
  \multicolumn{1}{c}{\textbf{Consistency}} &
  \multicolumn{1}{c}{\textbf{Usefulness}} &
  \multicolumn{1}{c}{\textbf{Clarification}} \\
  \midrule

{\textbf{\linuxflDiag}} &
  {2.5627} &
  {2.5986} &
  {4.1541}
\\ 
{\textbf{\agentlessDiag}} &
  {2.5699} &
  {2.6093} &
  {4.1111}
\\ 
\textbf{\toolname{}} &
  \textbf{2.9821} &
  \textbf{3.0789} &
  \textbf{4.7133} \\
\bottomrule
\end{tabular}
\vspace{-0.5em}
\end{table}

%% file: figures/failure_analysis.tex
\begin{figure}[t!]
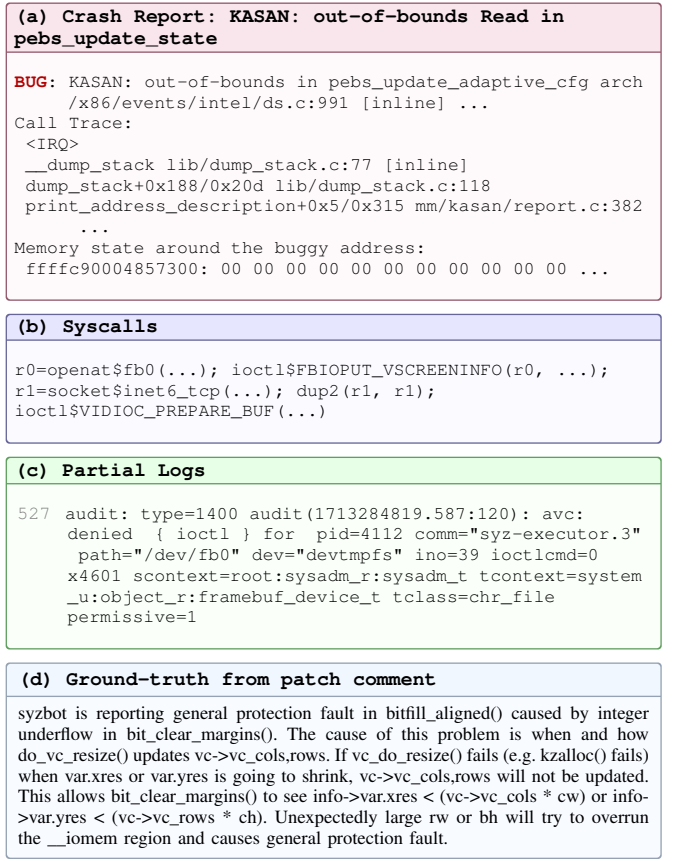

\centering
\begin{minipage}{0.98\columnwidth}
\begin{reportbox}{(a) Crash Report: KASAN: out-of-bounds Read in pebs\_update\_state}
BUG: KASAN: out-of-bounds in pebs_update_adaptive_cfg arch/x86/events/intel/ds.c:991 [inline] ...
Call Trace:
 <IRQ>
 __dump_stack lib/dump_stack.c:77 [inline]
 dump_stack+0x188/0x20d lib/dump_stack.c:118
 print_address_description+0x5/0x315 mm/kasan/report.c:382 ...
Memory state around the buggy address:
 ffffc90004857300: 00 00 00 00 00 00 00 00 00 00 00 ...
\end{reportbox}

\vspace{-2.5mm}
\begin{syscallbox}{(b) Syscalls}
r0=openat$fb0(...); ioctl$FBIOPUT_VSCREENINFO(r0, ...);
r1=socket$inet6_tcp(...); dup2(r1, r1); 
ioctl$VIDIOC_PREPARE_BUF(...)
\end{syscallbox}
\vspace{-2.5mm}
\begin{logbox}{(c) Partial Logs}
(*@\logln{527}@*)audit: type=1400 audit(1713284819.587:120): avc:  denied  { ioctl } for  pid=4112 comm="syz-executor.3" path="/dev/fb0" dev="devtmpfs" ino=39 ioctlcmd=0x4601 scontext=root:sysadm_r:sysadm_t tcontext=system_u:object_r:framebuf_device_t tclass=chr_file permissive=1
\end{logbox}
\vspace{-2.5mm}
\begin{textbox}{(d) Ground-truth from patch comment}
syzbot is reporting general protection fault in bitfill\_aligned() caused by integer underflow in bit\_clear\_margins(). The cause of this problem is when and how do\_vc\_resize() updates vc->vc\_{cols,rows}.
If vc\_do\_resize() fails (e.g. kzalloc() fails) when var.xres or var.yres is going to shrink, vc->vc\_{cols,rows} will not be updated. This allows bit\_clear\_margins() to see info->var.xres < (vc->vc\_cols * cw) or info->var.yres < (vc->vc\_rows * ch). Unexpectedly large rw or bh will try to overrun the \_\_iomem region and causes general protection fault.
\end{textbox}

\end{minipage}
\caption{A failure diagnosis by \toolname{}}
\label{fig:failure_diagnosis}
\vspace{-0.5em}
\end{figure}

%% file: scripts/7_related_work.tex
\section{Related Work}
\label{sec:related_work}

Root-cause diagnosis aims to identify the underlying cause of a system failure. Numerous studies have proposed automated diagnosis techniques for complex software systems, especially distributed systems and cloud service systems, based on diverse runtime data sources such as metrics, logs, and traces. 
Traditional approaches typically learn from historical cases by training machine learning or deep learning models to capture failure-related patterns and support tasks such as failure-type classification and faulty-service localization~\cite{soldani2022anomaly}. 
Some of these methods rely on a single type of diagnostic signal, such as logs~\cite{lin2016log,zawawy2010log}, traces~\cite{guo2020graph,li2021practical}, or metrics~\cite{li2022actionable,ma2020diagnosing}, whereas others~\cite{yu2023nezha,zheng2024mulan,wu2021microdiag,zhang2023robust,lee2023eadro,hou2021diagnosing} integrate multiple sources of runtime evidence to support more comprehensive diagnosis.

Recent work has introduced LLMs into RCA by leveraging their strengths in semantic understanding, reasoning, and explanation generation. 
Early studies mainly treated RCA as a text generation or classification task, where an LLM predicts the root-cause category, faulty component, or mitigation strategy from incident descriptions and historical cases~\cite{ahmed2023recommending,chen2024automatic,jiang2025l4,zhang2025scalalog}. 
More recent approaches further extend LLM-based RCA to retrieval-augmented and agent-based settings, allowing models to iteratively inspect diagnostic artifacts and refine hypotheses during analysis~\cite{yao2022react,roy2024exploring,wang2024rcagent,xu2025openrca}.
However, these approaches are designed primarily for AIOps, where rich operational context is typically available and diagnosis usually targets faulty services, failure categories, or mitigation actions rather than fine-grained root-cause localization and explanation. 
Such assumptions do not readily hold for Linux kernel crashes, limiting the applicability of these approaches in the kernel setting.
To address this gap, we propose \toolname{}, an agent-based framework for fine-grained Linux kernel root-cause diagnosis.

%% file: scripts/8_conclusion.tex
\section{Conclusion}

We propose \toolname{}, an agent-based framework for kernel root-cause diagnosis grounded in structured reasoning over heterogeneous evidence.
\toolname{} aligns runtime artifacts with code-level semantics and employs role-specialized agents to analyze syscalls, logs, and crash reports from complementary perspectives. 
During the reasoning process, each agent iteratively explores the kernel source code and environment artifacts while incrementally constructing a structured Evidence Graph that captures inferred fault-propagation relationships.
Extensive evaluation on real-world kernel crashes demonstrates that \toolname{} consistently outperforms existing approaches in root-cause localization.
Moreover, both human and LLM-assisted evaluations confirm that \toolname{} produces root-cause explanations that are accurate, coherent, and actionable for debugging. 
\toolname{} provides an effective solution for bridging the gap between failure symptoms and their underlying causes in complex kernel systems.

%% file: reference.bib
@article{li2025issue,
  title={Issue-oriented agent-based framework for automated review comment generation},
  author={Li, Shuochuan and Wang, Dong and Thongtanunam, Patanamon and Wang, Zan and Yu, Jiuqiao and Chen, Junjie},
  xjournal={ACM Transactions on Software Engineering and Methodology},
  journal={ACM TOSEM},
  year={2025},
  publisher={ACM New York, NY}
}

@inproceedings{li2025coca,
  title={COCA: Generative Root Cause Analysis for Distributed Systems with Code Knowledge},
  author={Li, Yichen and Wu, Yulun and Liu, Jinyang and Jiang, Zhihan and Chen, Zhuangbin and Yu, Guangba and Lyu, Michael R},
  xbooktitle={2025 IEEE/ACM 47th International Conference on Software Engineering (ICSE)},
  booktitle={Proceedings of ICSE},
  pages={1346--1358},
  year={2025},
  organization={IEEE}
}

@article{xin2023causalrca,
  title={Causalrca: Causal inference based precise fine-grained root cause localization for microservice applications},
  author={Xin, Ruyue and Chen, Peng and Zhao, Zhiming},
  journal={Journal of Systems and Software},
  volume={203},
  pages={111724},
  year={2023},
  publisher={Elsevier}
}

@inproceedings{li2022causal,
  title={Causal inference-based root cause analysis for online service systems with intervention recognition},
  author={Li, Mingjie and Li, Zeyan and Yin, Kanglin and Nie, Xiaohui and Zhang, Wenchi and Sui, Kaixin and Pei, Dan},
  xbooktitle={Proceedings of the 28th ACM SIGKDD conference on knowledge discovery and data mining},
  booktitle={Proceedings of ACM SIGKDD (KDD)},
  pages={3230--3240},
  year={2022}
}

@article{shu2025empirical,
  title={An Empirical Study on Language Models for Generating Log Statements in Test Code},
  author={Shu, Honglin and Wang, Dong and Mastropaolo, Antonio and Bavota, Gabriele and Kamei, Yasutaka},
  xjournal={ACM Transactions on Software Engineering and Methodology},
  journal={ACM TOSEM},
  year={2025},
  publisher={ACM New York, NY}
}

@inproceedings{ahmed2023recommending,
  title={Recommending root-cause and mitigation steps for cloud incidents using large language models},
  author={Ahmed, Toufique and Ghosh, Supriyo and Bansal, Chetan and Zimmermann, Thomas and Zhang, Xuchao and Rajmohan, Saravan},
  xbooktitle={2023 IEEE/ACM 45th International Conference on Software Engineering (ICSE)},
  booktitle={Proceedings of ICSE},
  pages={1737--1749},
  year={2023},
  organization={IEEE}
}

@inproceedings{tian2024large,
  title={Large language models for equivalent mutant detection: How far are we?},
  author={Tian, Zhao and Shu, Honglin and Wang, Dong and Cao, Xuejie and Kamei, Yasutaka and Chen, Junjie},
  xbooktitle={Proceedings of the 33rd ACM SIGSOFT International Symposium on Software Testing and Analysis},
  booktitle={Proceedings of ISSTA},
  pages={1733--1745},
  year={2024}
}

@inproceedings{chen2024automatic,
  title={Automatic root cause analysis via large language models for cloud incidents},
  xauthor={Chen, Yinfang and Xie, Huaibing and Ma, Minghua and Kang, Yu and Gao, Xin and Shi, Liu and Cao, Yunjie and Gao, Xuedong and Fan, Hao and Wen, Ming and others},
  author={Chen, Yinfang and Xie, Huaibing and Ma, Minghua and Kang, Yu and Gao, Xin and Shi, Liu and Cao, Yunjie and Gao, Xuedong and Fan, Hao and Wen, Ming and Zeng, Jun and Ghosh, Supriyo and Zhang, Xuchao and Zhang, Chaoyun and Lin, Qingwei and Rajmohan, Saravan and Zhang, Dongmei and Xu, Tianyin},
  xbooktitle={Proceedings of the Nineteenth European Conference on Computer Systems},
  booktitle = {Proceedings of EuroSys},
  pages={674--688},
  year={2024}
}

@inproceedings{yao2022react,
  title={React: Synergizing reasoning and acting in language models},
  author={Yao, Shunyu and Zhao, Jeffrey and Yu, Dian and Du, Nan and Shafran, Izhak and Narasimhan, Karthik R and Cao, Yuan},
  xbooktitle={The eleventh international conference on learning representations},
  booktitle = {Proceedings of ICLR},
  year={2022}
}

@inproceedings{roy2024exploring,
  title={Exploring llm-based agents for root cause analysis},
  author={Roy, Devjeet and Zhang, Xuchao and Bhave, Rashi and Bansal, Chetan and Las-Casas, Pedro and Fonseca, Rodrigo and Rajmohan, Saravan},
  xbooktitle={Companion proceedings of the 32nd ACM international conference on the foundations of software engineering},
  booktitle={Proceedings of FSE}, 
  pages={208--219},
  year={2024}
}

@inproceedings{wang2024rcagent,
  title={Rcagent: Cloud root cause analysis by autonomous agents with tool-augmented large language models},
  author={Wang, Zefan and Liu, Zichuan and Zhang, Yingying and Zhong, Aoxiao and Wang, Jihong and Yin, Fengbin and Fan, Lunting and Wu, Lingfei and Wen, Qingsong},
  xbooktitle={Proceedings of the 33rd ACM international conference on information and knowledge management},
  booktitle = {Proceedings of CIKM},
  pages={4966--4974},
  year={2024}
}

@inproceedings{xu2025openrca,
  title={Openrca: Can large language models locate the root cause of software failures?},
  author={Xu, Junjielong and Zhang, Qinan and Zhong, Zhiqing and He, Shilin and Zhang, Chaoyun and Lin, Qingwei and Pei, Dan and He, Pinjia and Zhang, Dongmei and Zhang, Qi},
  xbooktitle={The thirteenth international conference on learning representations},
  xxbooktitle = {International Conference on Learning Representations (ICLR)},
  booktitle = {Proceedings of ICLR},
  year={2025}
}

@article{soldani2022anomaly,
  title={Anomaly detection and failure root cause analysis in (micro) service-based cloud applications: A survey},
  author={Soldani, Jacopo and Brogi, Antonio},
  journal={ACM Computing Surveys (CSUR)},
  volume={55},
  number={3},
  pages={1--39},
  year={2022},
  publisher={ACM New York, NY}
}

@inproceedings{jiang2025l4,
  title={L4: Diagnosing large-scale llm training failures via automated log analysis},
  author={Jiang, Zhihan and Huang, Junjie and Yu, Guangba and Chen, Zhuangbin and Li, Yichen and Zhong, Renyi and Feng, Cong and Yang, Yongqiang and Yang, Zengyin and Lyu, Michael},
  xbooktitle={Proceedings of the 33rd ACM International Conference on the Foundations of Software Engineering},
  booktitle={Proceedings of FSE},
  pages={51--63},
  year={2025}
}

@inproceedings{zhang2025scalalog,
  title={Scalalog: Scalable log-based failure diagnosis using llm},
  author={Zhang, Lingzhe and Jia, Tong and Jia, Mengxi and Wu, Yifan and Liu, Hongyi and Li, Ying},
  xbooktitle={ICASSP 2025-2025 IEEE International Conference on Acoustics, Speech and Signal Processing (ICASSP)},
  booktitle={Proceedings of ICASSP},
  pages={1--5},
  year={2025},
  organization={IEEE}
}

@misc{linux_kernel_40m,
  author       = {{Stackscale}},
  title        = {Linux Kernel Surpasses 40 Million Lines of Code},
  year         = {2024},
  howpublished = {\url{https://www.stackscale.com/blog/linux-kernel-surpasses-40-million-lines-code}},
  note         = {Accessed: 2026-03-18}
}

@misc{syzbot,
  title = {Syzbot},
  howpublished = {\url{https://syzkaller.appspot.com}},
  year={2025}
}

@misc{syzkaller,
  author       = {Google},
  title        = {syzkaller: an unsupervised coverage‑guided kernel fuzzer},
  howpublished = {\url{https://github.com/google/syzkaller}},
  year         = {2015},
  note         = {Accessed: 2026-03-18}
}

@inproceedings{zhou2025benchmarking,
  title={Taming System Complexity: Demystifying Software Engineering Agents in Diagnosing Linux Kernel Faults},
  author={Zhou, Zhenhao and Huang, Zhuochen and He, Yike and Wang, Chong and Wang, Jiajun and Wu, Yijian and Peng, Xin and Lou, Yiling},
  xbooktitle={Proceedings of the 64th Annual Meeting of the Association for Computational Linguistics (Volume 1: Long Papers)},
  booktitle={Proceedings of ACL},
  pages={18899--18916},
  year={2026}
}

@article{mathai2024kgym,
  title={Kgym: A platform and dataset to benchmark large language models on linux kernel crash resolution},
  author={Mathai, Alex and Huang, Chenxi and Maniatis, Petros and Nogikh, Aleksandr and Ivan{\v{c}}i{\'c}, Franjo and Yang, Junfeng and Ray, Baishakhi},
  journal={Advances in Neural Information Processing Systems},
  volume={37},
  pages={78053--78078},
  year={2024}
}

@article{yang2024swe,
  title={Swe-agent: Agent-computer interfaces enable automated software engineering},
  author={Yang, John and Jimenez, Carlos E and Wettig, Alexander and Lieret, Kilian and Yao, Shunyu and Narasimhan, Karthik and Press, Ofir},
  journal={Advances in Neural Information Processing Systems},
  volume={37},
  pages={50528--50652},
  year={2024}
}

@article{xia2024agentless,
  title={Demystifying llm-based software engineering agents},
  author={Xia, Chunqiu Steven and Deng, Yinlin and Dunn, Soren and Zhang, Lingming},
  journal={Proceedings of the ACM on Software Engineering},
  volume={2},
  number={FSE},
  pages={801--824},
  year={2025},
  publisher={ACM New York, NY, USA}
}

@article{jiang2025deep,
  title={Deep assessment of code review generation approaches: Beyond lexical similarity},
  author={Jiang, Yanjie and Liu, Hui and Chen, Tianyi and Fan, Fu and Dong, Chunhao and Liu, Kui and Zhang, Lu},
  journal={arXiv preprint arXiv:2501.05176},
  year={2025}
}

@inproceedings{lu2025deepcrceval,
  title={Deepcrceval: Revisiting the evaluation of code review comment generation},
  author={Lu, Junyi and Li, Xiaojia and Hua, Zihan and Yu, Lei and Cheng, Shiqi and Yang, Li and Zhang, Fengjun and Zuo, Chun},
  booktitle={International Conference on Fundamental Approaches to Software Engineering},
  pages={43--64},
  year={2025},
  organization={Springer}
}

@inproceedings{sun2025bitsai,
  title={Bitsai-cr: Automated code review via llm in practice},
  author={Sun, Tao and Xu, Jian and Li, Yuanpeng and Yan, Zhao and Zhang, Ge and Xie, Lintao and Geng, Lu and Wang, Zheng and Chen, Yueyan and Lin, Qin and others},
  xbooktitle={Proceedings of the 33rd ACM International Conference on the Foundations of Software Engineering},
  booktitle={Proceedings of FSE},
  pages={274--285},
  year={2025}
}

@article{tufano2024code,
  title={Code review automation: strengths and weaknesses of the state of the art},
  author={Tufano, Rosalia and Dabi{\'c}, Ozren and Mastropaolo, Antonio and Ciniselli, Matteo and Bavota, Gabriele},
  xjournal={IEEE Transactions on Software Engineering},
  journal={IEEE TSE},
  volume={50},
  number={2},
  pages={338--353},
  year={2024},
  publisher={IEEE}
}

@article{cohen1968weighted,
  title={Weighted kappa: Nominal scale agreement provision for scaled disagreement or partial credit.},
  author={Cohen, Jacob},
  journal={Psychological bulletin},
  volume={70},
  number={4},
  pages={213},
  year={1968},
  publisher={American Psychological Association}
}

@inproceedings{li2021fault,
  title     = {Fault localization with code coverage representation learning},
  author    = {Li, Y. and Wang, S. and Nguyen, T.},
  xbooktitle = {Proceedings of the 2021 IEEE/ACM 43rd International Conference on Software Engineering (ICSE)},
  booktitle = {Proceedings of ICSE},
  pages     = {661--673},
  year      = {2021},
  organization = {IEEE}
}

@inproceedings{li2019deepfl,
  title={DeepFL: Integrating Multiple Fault Diagnosis Dimensions for Deep Fault Localization},
  author={Li, Xia and Li, Wei and Zhang, Yuqun and Zhang, Lingming},
  xbooktitle={Proceedings of the 28th ACM SIGSOFT International Symposium on Software Testing and Analysis (ISSTA '19)},
  booktitle={Proceedings of ISSTA},
  pages={169--180},
  year={2019},
  organization={ACM}
}

@article{zou2019empirical,
  title={An empirical study of fault localization families and their combinations},
  author={Zou, Daming and Liang, Jingjing and Xiong, Yingfei and Ernst, Michael D and Zhang, Lu},
  xjournal={IEEE Transactions on Software Engineering},
  journal={IEEE TSE},
  volume={47},
  number={2},
  pages={332--347},
  year={2019},
  publisher={IEEE}
}

@article{dem2006statistical,
  title={Statistical Comparisons of Classifiers over Multiple Data Sets},
  author={Demšar, Janez},
  journal={Journal of Machine Learning Research},
  volume={7},
  pages={1--30},
  year={2006}
}

@article{liu2024deepseek,
  title={Deepseek-v3 technical report},
  author={Liu, Aixin and Feng, Bei and Xue, Bing and Wang, Bingxuan and Wu, Bochao and Lu, Chengda and Zhao, Chenggang and Deng, Chengqi and Zhang, Chenyu and Ruan, Chong and others},
  journal={arXiv preprint arXiv:2412.19437},
  year={2024}
}

@misc{linux_kernel_org,
  author       = {{The Linux Kernel Organization}},
  title        = {{The Linux Kernel Archives}},
  year         = {2026},
  howpublished = {\url{https://www.kernel.org/}},
  note         = {Accessed: 2026-03-26}
}

@article{xia2023information,
  title={Information Retrieval-Based Techniques for Software Fault Localization},
  author={Xia, Xin and Lo, David},
  journal={Handbook of Software Fault Localization: Foundations and Advances},
  pages={365--391},
  year={2023},
  publisher={Wiley Online Library}
}

@article{zhang2019finelocator,
  title={FineLocator: A novel approach to method-level fine-grained bug localization by query expansion},
  author={Zhang, Wen and Li, Ziqiang and Wang, Qing and Li, Juan},
  journal={Information and Software Technology},
  volume={110},
  pages={121--135},
  year={2019},
  publisher={Elsevier}
}

@article{chen2021pathidea,
  title={Pathidea: Improving information retrieval-based bug localization by re-constructing execution paths using logs},
  author={Chen, An Ran and Chen, Tse-Hsun and Wang, Shaowei},
  xjournal={IEEE Transactions on Software Engineering},
  journal={IEEE TSE},
  volume={48},
  number={8},
  pages={2905--2919},
  year={2021},
  publisher={IEEE}
}

@inproceedings{razzaq2021boostnsift,
  title={BoostNSift: A query boosting and code sifting technique for method level bug localization},
  author={Razzaq, Abdul and Buckley, Jim and Patten, James Vincent and Chochlov, Muslim and Sai, Ashish Rajendra},
  booktitle={2021 IEEE 21st International Working Conference on Source Code Analysis and Manipulation (SCAM)},
  pages={81--91},
  year={2021},
  organization={IEEE}
}

@inproceedings{papineni2002bleu,
  title={Bleu: a method for automatic evaluation of machine translation},
  author={Papineni, Kishore and Roukos, Salim and Ward, Todd and Zhu, Wei-Jing},
  xbooktitle={Proceedings of the 40th annual meeting of the Association for Computational Linguistics},
  booktitle={Proceedings of ACL},
  pages={311--318},
  year={2002}
}

@inproceedings{lin2004rouge,
  title={Rouge: A package for automatic evaluation of summaries},
  author={Lin, Chin-Yew},
  booktitle={Text summarization branches out},
  pages={74--81},
  year={2004}
}

@inproceedings{li2021practical,
  title={Practical root cause localization for microservice systems via trace analysis},
  author={Li, Zeyan and Chen, Junjie and Jiao, Rui and Zhao, Nengwen and Wang, Zhijun and Zhang, Shuwei and Wu, Yanjun and Jiang, Long and Yan, Leiqin and Wang, Zikai and others},
  xbooktitle={2021 IEEE/ACM 29th International Symposium on Quality of Service (IWQOS)},
  booktitle={Proceedings of IWQOS},
  pages={1--10},
  year={2021},
  organization={IEEE}
}

@inproceedings{liu2021microhecl,
  title={Microhecl: High-efficient root cause localization in large-scale microservice systems},
  author={Liu, Dewei and He, Chuan and Peng, Xin and Lin, Fan and Zhang, Chenxi and Gong, Shengfang and Li, Ziang and Ou, Jiayu and Wu, Zheshun},
  xbooktitle={2021 IEEE/ACM 43rd International Conference on Software Engineering: Software Engineering in Practice (ICSE-SEIP)},
  booktitle={Proceedings of ICSE-SEIP},
  pages={338--347},
  year={2021},
  organization={IEEE}
}

@inproceedings{xie2023unsupervised,
  title={Unsupervised anomaly detection on microservice traces through graph vae},
  author={Xie, Zhe and Xu, Haowen and Chen, Wenxiao and Li, Wanxue and Jiang, Huai and Su, Liangfei and Wang, Hanzhang and Pei, Dan},
  xbooktitle={Proceedings of the ACM Web Conference 2023},
  booktitle={Proceedings of the ACM Web Conference},
  pages={2874--2884},
  year={2023}
}

@inproceedings{zeng2023traceark,
  title={Traceark: Towards actionable performance anomaly alerting for online service systems},
  author={Zeng, Zhengran and Zhang, Yuqun and Xu, Yong and Ma, Minghua and Qiao, Bo and Zou, Wentao and Chen, Qingjun and Zhang, Meng and Zhang, Xu and Zhang, Hongyu and others},
  xbooktitle={2023 IEEE/ACM 45th International Conference on Software Engineering: Software Engineering in Practice (ICSE-SEIP)},
  booktitle={Proceedings of ICSE-SEIP},
  pages={258--269},
  year={2023},
  organization={IEEE}
}

@article{sui2023logkg,
  title={Logkg: Log failure diagnosis through knowledge graph},
  author={Sui, Yicheng and Zhang, Yuzhe and Sun, Jianjun and Xu, Ting and Zhang, Shenglin and Li, Zhengdan and Sun, Yongqian and Guo, Fangrui and Shen, Junyu and Zhang, Yuzhi and others},
  journal={IEEE Transactions on Services Computing},
  volume={16},
  number={5},
  pages={3493--3507},
  year={2023},
  publisher={IEEE}
}

@inproceedings{zhang2024multivariate,
  title={Multivariate log-based anomaly detection for distributed database},
  author={Zhang, Lingzhe and Jia, Tong and Jia, Mengxi and Li, Ying and Yang, Yong and Wu, Zhonghai},
  xbooktitle={Proceedings of the 30th ACM SIGKDD Conference on Knowledge Discovery and Data Mining},
  booktitle={Proceedings of ACM SIGKDD (KDD)},
  pages={4256--4267},
  year={2024}
}

@article{zhang2023robust,
title={Robust Failure Diagnosis of Microservice System through Multimodal Data},
author={Zhang, Shenglin and Jin, Pengxiang and Lin, Zihan and Sun, Yongqian and Zhang, Bicheng and Xia, Sibo and Li, Zhengdan and Zhong, Zhenyu and Ma, Minghua and Jin, Wa and others},
journal={IEEE Transactions on Services Computing},
number={01},
pages={1--14},
year={2023},
publisher={IEEE Computer Society}
}

@inproceedings{parfenova2024automating,
  title={Automating qualitative data analysis with large language models},
  author={Parfenova, Angelina and Denzler, Alexander and Pfeffer, J{\"o}rgen},
  xbooktitle={Proceedings of the 62nd Annual Meeting of the Association for Computational Linguistics (Volume 4: Student Research Workshop)},
  booktitle={Proceedings of ACL},
  pages={83--91},
  year={2024}
}

@book{krippendorff2018content,
  title={Content analysis: An introduction to its methodology},
  author={Krippendorff, Klaus},
  year={2018},
  publisher={Sage publications}
}

@inproceedings{hartvigsen2022toxigen,
  title={Toxigen: A large-scale machine-generated dataset for adversarial and implicit hate speech detection},
  author={Hartvigsen, Thomas and Gabriel, Saadia and Palangi, Hamid and Sap, Maarten and Ray, Dipankar and Kamar, Ece},
  xbooktitle={Proceedings of the 60th annual meeting of the association for computational linguistics (volume 1: Long papers)},
  booktitle={Proceedings of ACL},
  pages={3309--3326},
  year={2022}
}

@article{li2025yesterday,
  title={Yesterday once more: facilitating Linux kernel bug reproduction via reverse fuzzing},
  author={Li, Xingwei and Kang, Yan and Wu, Chenggang and Liu, Danjun and Wang, Jiming and Sun, Yue and Wu, Zehui and Wang, Yunchao and Ma, Rongkuan and Wei, Qiang},
  xjournal={IEEE Transactions on Information Forensics and Security},
  journal={IEEE TIFS},
  year={2025},
  publisher={IEEE}
}

@inproceedings{lin2022grebe,
  title={Grebe: Unveiling exploitation potential for linux kernel bugs},
  author={Lin, Zhenpeng and Chen, Yueqi and Wu, Yuhang and Mu, Dongliang and Yu, Chensheng and Xing, Xinyu and Li, Kang},
  booktitle={2022 IEEE Symposium on Security and Privacy (SP)},
  pages={2078--2095},
  year={2022},
  organization={IEEE}
}

@inproceedings{lin2016log,
  title={Log clustering based problem identification for online service systems},
  author={Lin, Qingwei and Zhang, Hongyu and Lou, Jian-Guang and Zhang, Yu and Chen, Xuewei},
  xbooktitle={Proceedings of the 38th international conference on software engineering companion},
  booktitle={Proceedings of ICSE},
  pages={102--111},
  year={2016}
}

@inproceedings{yu2023nezha,
  title={Nezha: Interpretable fine-grained root causes analysis for microservices on multi-modal observability data},
  author={Yu, Guangba and Chen, Pengfei and Li, Yufeng and Chen, Hongyang and Li, Xiaoyun and Zheng, Zibin},
  xbooktitle={Proceedings of the 31st ACM joint European software engineering conference and symposium on the foundations of software engineering},
  booktitle={Proceedings of ESEC/FSE},
  pages={553--565},
  year={2023}
}

@inproceedings{guo2020graph,
  title={Graph-based trace analysis for microservice architecture understanding and problem diagnosis},
  author={Guo, Xiaofeng and Peng, Xin and Wang, Hanzhang and Li, Wanxue and Jiang, Huai and Ding, Dan and Xie, Tao and Su, Liangfei},
  xbooktitle={Proceedings of the 28th ACM joint meeting on European software engineering conference and symposium on the foundations of software engineering},
  booktitle={Proceedings of ESEC/FSE},
  pages={1387--1397},
  year={2020}
}

@inproceedings{zheng2024mulan,
  title={Mulan: Multi-modal causal structure learning and root cause analysis for microservice systems},
  author={Zheng, Lecheng and Chen, Zhengzhang and He, Jingrui and Chen, Haifeng},
  booktitle={Proceedings of the ACM Web Conference 2024},
  pages={4107--4116},
  year={2024}
}

@inproceedings{wu2021microdiag,
  title={Microdiag: Fine-grained performance diagnosis for microservice systems},
  author={Wu, Li and Tordsson, Johan and Bogatinovski, Jasmin and Elmroth, Erik and Kao, Odej},
  booktitle={2021 IEEE/ACM International Workshop on Cloud Intelligence (CloudIntelligence)},
  pages={31--36},
  year={2021},
  organization={IEEE}
}

@inproceedings{lee2023eadro,
  title={Eadro: An end-to-end troubleshooting framework for microservices on multi-source data},
  author={Lee, Cheryl and Yang, Tianyi and Chen, Zhuangbin and Su, Yuxin and Lyu, Michael R},
  xbooktitle={2023 IEEE/ACM 45th International Conference on Software Engineering (ICSE)},
  booktitle={Proceedings of ICSE}, 
  pages={1750--1762},
  year={2023},
  organization={IEEE}
}

@inproceedings{huo2023autolog,
  title={Autolog: A log sequence synthesis framework for anomaly detection},
  author={Huo, Yintong and Li, Yichen and Su, Yuxin and He, Pinjia and Xie, Zifan and Lyu, Michael R},
  xbooktitle={2023 38th IEEE/ACM International Conference on Automated Software Engineering (ASE)},
  booktitle={Proceedings of ASE}, 
  pages={497--509},
  year={2023},
  organization={IEEE}
}

@article{patel2022sense,
  title={The sense of logging in the linux kernel},
  author={Patel, Keyur and Faccin, Jo{\~a}o and Hamou-Lhadj, Abdelwahab and Nunes, Ingrid},
  journal={Empirical Software Engineering},
  volume={27},
  number={6},
  pages={153},
  year={2022},
  publisher={Springer}
}

@inproceedings{hou2021diagnosing,
  title={Diagnosing performance issues in microservices with heterogeneous data source},
  author={Hou, Chuanjia and Jia, Tong and Wu, Yifan and Li, Ying and Han, Jing},
  xbooktitle={2021 IEEE Intl Conf on Parallel \& Distributed Processing with Applications, Big Data \& Cloud Computing, Sustainable Computing \& Communications, Social Computing \& Networking (ISPA/BDCloud/SocialCom/SustainCom)},
  booktitle = {Proceedings of ISPA/BDCloud/SocialCom/SustainCom},
  pages={493--500},
  year={2021},
  organization={IEEE}
}

@inproceedings{li2022actionable,
  title={Actionable and interpretable fault localization for recurring failures in online service systems},
  author={Li, Zeyan and Zhao, Nengwen and Li, Mingjie and Lu, Xianglin and Wang, Lixin and Chang, Dongdong and Nie, Xiaohui and Cao, Li and Zhang, Wenchi and Sui, Kaixin and others},
  xbooktitle={Proceedings of the 30th ACM Joint European Software Engineering Conference and Symposium on the Foundations of Software Engineering},
  booktitle={Proceedings of ESEC/FSE},
  pages={996--1008},
  year={2022}
}

@article{ma2020diagnosing,
  title={Diagnosing root causes of intermittent slow queries in cloud databases},
  author={Ma, Minghua and Yin, Zheng and Zhang, Shenglin and Wang, Sheng and Zheng, Christopher and Jiang, Xinhao and Hu, Hanwen and Luo, Cheng and Li, Yilin and Qiu, Nengjun and others},
  journal={Proceedings of the VLDB Endowment},
  volume={13},
  number={8},
  pages={1176--1189},
  year={2020},
  publisher={VLDB Endowment}
}

@inproceedings{zawawy2010log,
  title={Log filtering and interpretation for root cause analysis},
  author={Zawawy, Hamzeh and Kontogiannis, Kostas and Mylopoulos, John},
  booktitle={2010 IEEE International Conference on Software Maintenance},
  pages={1--5},
  year={2010},
  organization={IEEE}
}

@inproceedings{chen2020towards,
  title={Towards intelligent incident management: why we need it and how we make it},
  author={Chen, Zhuangbin and Kang, Yu and Li, Liqun and Zhang, Xu and Zhang, Hongyu and Xu, Hui and Zhou, Yangfan and Yang, Li and Sun, Jeffrey and Xu, Zhangwei and others},
  xbooktitle={Proceedings of the 28th ACM Joint Meeting on European Software Engineering Conference and Symposium on the Foundations of Software Engineering},
  booktitle={Proceedings of ESEC/FSE},
  pages={1487--1497},
  year={2020}
}

@inproceedings{hao2023syzdescribe,
  title={Syzdescribe: Principled, automated, static generation of syscall descriptions for kernel drivers},
  author={Hao, Yu and Li, Guoren and Zou, Xiaochen and Chen, Weiteng and Zhu, Shitong and Qian, Zhiyun and Sani, Ardalan Amiri},
  xbooktitle={2023 IEEE Symposium on Security and Privacy (SP)},
  booktitle={IEEE S\&P},
  pages={3262--3278},
  year={2023},
  organization={IEEE}
}

@article{wilcoxon1945individual,
  title={Individual comparisons by ranking methods},
  author={Wilcoxon, Frank},
  journal={Biometrics bulletin},
  volume={1},
  number={6},
  pages={80--83},
  year={1945},
  publisher={JSTOR}
}

@inproceedings{
    jimenez2024swebench,
    title={{SWE}-bench: Can Language Models Resolve Real-world Github Issues?},
    author={Carlos E Jimenez and John Yang and Alexander Wettig and Shunyu Yao and Kexin Pei and Ofir Press and Karthik R Narasimhan},
    xbooktitle={The Twelfth International Conference on Learning Representations},
    booktitle={Proceedings of ICLR},
    year={2024},
}

@article{wang2025empirical,
  title={An empirical study of test case prioritization on the Linux Kernel},
  author={Wang, Haichi and Yu, Ruiguo and Wang, Dong and Du, Yiheng and Zhao, Yingquan and Chen, Junjie and Wang, Zan},
  journal={Automated Software Engineering},
  volume={32},
  number={2},
  pages={49},
  year={2025},
  publisher={Springer}
}

@misc{kerneldiag2026,
    title = {{KernelDiag-Artifact}},
    url = {https://github.com/VikingStudyHard/KernelDiag-Artifact},
    author = {},
    year = {Accessed: 2026}
}

@inproceedings{chen2025locagent,

  title={Locagent: Graph-guided llm agents for code localization},
  author={Chen, Zhaoling and Tang, Robert and Deng, Gangda and Wu, Fang and Wu, Jialong and Jiang, Zhiwei and Prasanna, Viktor and Cohan, Arman and Wang, Xingyao},
  xbooktitle={Proceedings of the 63rd Annual Meeting of the Association for Computational Linguistics (volume 1: Long papers)},
  booktitle={Proceedings of ACL},
  pages={8697--8727},
  year={2025}
}

@article{wen2019historical,
  title={Historical spectrum based fault localization},
  author={Wen, Ming and Chen, Junjie and Tian, Yongqiang and Wu, Rongxin and Hao, Dan and Han, Shi and Cheung, Shing-Chi},
  xjournal={IEEE Transactions on Software Engineering},
  journal={IEEE TSE},
  volume={47},
  number={11},
  pages={2348--2368},
  year={2019},
  publisher={IEEE}
}

@article{jia2010analysis,
  title={An analysis and survey of the development of mutation testing},
  author={Jia, Yue and Harman, Mark},
  xjournal={IEEE transactions on software engineering},
  journal={IEEE TSE},
  volume={37},
  number={5},
  pages={649--678},
  year={2010},
  publisher={IEEE}
}

@article{andrews2006using,
  title={Using mutation analysis for assessing and comparing testing coverage criteria},
  author={Andrews, James H and Briand, Lionel C and Labiche, Yvan and Namin, Akbar Siami},
  xjournal={IEEE Transactions on Software Engineering},
  journal={IEEE TSE},
  volume={32},
  number={8},
  pages={608--624},
  year={2006},
  publisher={IEEE}
}

@misc{qwen3max,
    title = {Qwen3-Max: Just Scale it},
    author = {Qwen Team},
    month = {September},
    year = {2025}
}

@article{mann1947test,
  title={On a test of whether one of two random variables is stochastically larger than the other},
  author={Mann, Henry B and Whitney, Donald R},
  journal={The annals of mathematical statistics},
  pages={50--60},
  year={1947},
  publisher={JSTOR}
}

@article{holm1979simple,
  title={A simple sequentially rejective multiple test procedure},
  author={Holm, Sture},
  journal={Scandinavian journal of statistics},
  pages={65--70},
  year={1979},
  publisher={JSTOR}
}

@article{cliff1993dominance,
  title={Dominance statistics: Ordinal analyses to answer ordinal questions.},
  author={Cliff, Norman},
  journal={Psychological bulletin},
  volume={114},
  number={3},
  pages={494},
  year={1993},
  publisher={American Psychological Association}
}
